\documentclass[aps,pra,reprint,floatfix]{revtex4-1} 
\usepackage[utf8]{inputenc}

\usepackage{natbib}
\usepackage{amsmath}
\usepackage{graphicx}
\usepackage{float}
\usepackage{physics}
\usepackage{amssymb}
\usepackage{appendix}

\usepackage[normalem]{ulem}
\newcommand{\stkout}[1]{\ifmmode\text{\sout{\ensuremath{#1}}}\else\sout{#1}\fi}

\usepackage{xcolor}

\begin{document}

\title{Fabry-Perot microcavity spectra have a fine structure}
\author{M. P. van Exter$^1$}
\author{M. Wubs$^2$}
\author{E. Hissink$^1$}
\author{C. Koks$^1$}
\affiliation{1. Huygens-Kamerlingh Onnes Laboratory, Leiden University,
P.O. Box 9504, 2300 RA Leiden, The Netherlands}
\affiliation{2. Department of Photonics Engineering, Technical University of Denmark, Ørsteds Plads 345A, DK-2800 Kgs. Lyngby, Denmark}

\begin{abstract} 
Optical cavities can support many transverse and longitudinal modes. 
A paraxial scalar theory predicts that the resonance frequencies of these modes cluster in different orders.
A non-paraxial vector theory predicts that the frequency degeneracy within these clusters is lifted, such that each order acquires a spectral fine structure, comparable to the fine structure observed in atomic spectra.
In this paper, we calculate this fine structure for microcavities and show how it originates from various non-paraxial effects and is co-determined by mirror aberrations. 
The presented theory, which applies perturbation theory to Maxwell's equations with boundary conditions, proves to be very powerful.
It generalizes the effective 1-dimensional description of Fabry-Perot cavities to a 3-dimensional multi-transverse-mode description. 
It thereby provides new physical insights in several mode-shaping effects and a detailed prediction of the fine structure in Fabry-Perot spectra. 
\end{abstract}

\maketitle

\section{Introduction}

Tunable Fabry-Perot (FP) cavities are popular tools in optics, where they are used as spectrum analyzer~\cite{Siegman} and as a means to resonantly trap light between two high-reflecting mirrors~\cite{Vahala2003}. 
An optical microcavity is a miniature version of a FP cavity, where the two mirrors are now positioned at only a few wavelengths $\lambda$ from each other and at least one of the mirrors has a radius of curvature $R_m \ll 100 \lambda$. 
Microcavities can strongly confine the optical field, boost the light-matter interaction of intra-cavity emitters~\cite{Najer2019,Vallance2016,Wang2019,Vogl2019, Haussler2021, Flatten2018}, and increase the collection efficiency and emitted fraction into the zero-phonon line~\cite{Ruf2021}.

Microcavities support compact optical modes with large opening angles. 
This can push their operation beyond the paraxial regime and can require a non-paraxial description of the optical propagation and a more thorough description of the mirror reflections. 
Elements of the resulting spectral finestructure have been reported \cite{Riedel2020, Trichet2018,Haussler2019,papageorge2016,Mader2015}, but a complete description was missing.

Non-paraxial corrections to the optical propagation were already analyzed in the seventies and eighties. 
Lax~\cite{Lax1975} describes a general framework that treats these corrections as different terms in a Taylor expansion. 
Erickson~\cite{Erickson,Erickson1977} calculates the scalar non-paraxial correction to the cavity resonances. 
Cullen~\cite{Cullen} and Davis~\cite{Davis1984} add a vector-correction to this description.
Yu and Luk~\cite{Yu1983,Yu1984} and Luk~\cite{Luk1986} are the first to combine these corrections in a complete analysis of the optical resonances in cavities with spherical mirrors. 
More recently, Zeppenfeld and Pinkse~\cite{Zeppenfeld2010} performed an alternative complete analysis of rotationally-symmetric cavities, using spheroidal wave functions.

Additional corrections occur when the mirrors are not spherical, but have astigmatic or more general deformations, common to microcavities \cite{Trichet2018,Benedikter2015,Benedikter2019}. 
Kleckner et al.~\cite{Kleckner2010} present a general framework to describe the effect of these deformations, but their description does not yield analytic solutions. 

This paper presents a general theoretical framework for the optical resonances in tunable FP microcavities.
The description is semi-analytic and exact in the limit of small perturbations.
The description uses a roundtrip operator that acts on field profiles and uses perturbation theory to calculate the effects of several deviations from the standard paraxial theory with spherical mirrors, including non-paraxial effects and deviations from the spherical mirror shape.
It thereby extends the standard 1-dimensional description of the Fabry-Perot interferometer to a  3-dimensional multi-transverse-mode description.
The presented mathematical and physical tools can be applied to a wide range of optical systems. 

The paper then applies this theory to calculate the fine structure in Fabry-Perot spectra. 
The paraxial scalar theory predicts that modes with the same longitudinal mode number $q$ and transverse order $N$ should be frequency degenerate.
But a more complete theory shows that each $(q,N)$-group exhibits a spectral fine structure, where modes with different radial mode number $p$, orbital angular momentum (OAM) mode number $\ell$, and polarization $v$ have slightly different resonance frequencies, even when they belong to the same $(q,N)$ order.
The paper analyses and classifies the different effects that contribute to this optical fine structure and identifies which ones are relevant under which conditions. 
It thus aims to present a complete description of this intriguing phenomenon. 

The fine structure in FP spectra is analogous to the fine structure observed in atomic spectra. 
For atomic spectra, the simple Bohr model of hydrogenic atoms predicts that their energy levels should only depend on the principal quantum number $n$.
But a more complete description, that among others includes relativistic corrections and spin-orbit coupling ~\cite{Griffiths}, shows that levels with the same $n$ are frequency split in the so-called fine structure with additional quantum numbers, $\ell$ for the orbital angular momentum and $s$ for the spin. 

The presented theory is inspired by our own experimental observations of intriguing fine structures in transmission spectra of FP microcavities \cite{Koks2022b} and the lack of an adequate theory.
Similar structures must have been observed by other groups, but hardly anything has been published.
For microwave cavities, Erickson~\cite{Erickson} measured the non-paraxial frequency splittings between radial modes and Yu and Luk~\cite{Yu1983} measured it between OAM modes. 
In the optical domain, Dufferwiel et al.~\cite{Dufferwiel} reported a fine structure in the $N=1$ group of an optical cavity filled with semiconductor quantum wells, but this TE-TM splitting was mainly due to polariton effects.
Zeppenfeld et al. reported an observation of a spectral fine structure at the CLEO 2010 conference~\cite{Zeppenfeld-CLEO}.
Raman spectra of Riedel et al.~\cite{Riedel2020} showed an astigmatic splitting, but did not show an additional fine structure, presumably because their cavities had a modest finesse $F$;
observable non-paraxial fine structures require $F\lambda/R_m \gtrsim 10$, for plano-concave cavities with mirror radius $R_m$ and wavelength $\lambda$. 
When more detailed experimental spectra become available, the most challenging aspect of their analysis will undoubtedly be the need to separate the, more fundamental, non-paraxial effects from the, more practical, mirror-shape effects. 
This paper describes how this can be done. 
Furthermore, we have already used the theory to analyze our own experiments on optical microcavities.

This paper consists of three parts: Part~I, in Secs.~\ref{sec:paraxial} and~\ref{sec:Roundtrip}, contains the description of the general framework, and ends with a preview of the key results in a table and a figure;  Part~II, in Secs.~\ref{sec:scalar} and~\ref{sec:vector}, presents the derivations of the fine structure for cavities with spherical mirrors. This part ends with a key equation that combines all relevant effects for plano-concave cavities, plus a comparison to the literature; Part~III, in Secs.~\ref{sec:Bragg} and~\ref{sec:mirror}, contains the analysis of several new effects for more general cavities. These new effects are additional scalar and vector corrections, including two astigmatic corrections.
Section~\ref{sec:Discussion} discusses the obtained results 
and addresses the potential role of residual corrections.
Section~\ref{sec:Summary} presents a summary and outlook. There are three Appendices.


\section{Paraxial scalar modes}
\label{sec:paraxial}

We consider the propagation and reflection of light in a plano-concave Fabry-Perot (FP) cavity.
Both mirrors are highly reflective and large enough to avoid clipping losses, such that all relevant modes are virtually lossless and spectrally well resolved. 
This cavity exhibits sharp resonances, visible as peaks in the optical transmission and dips in the optical reflection, at particular combinations of cavity length $L$ and optical wavelength $\lambda$, where light is resonantly trapped in the cavity. 
We want to derive the exact resonance conditions and spatial profiles of the associated eigenmodes.
This problem sounds simple but is surprisingly difficult; there are no exact solutions for the general non-paraxial case. 

The propagation of light in the paraxial limit is standard material in many textbooks on optics~\cite{Siegman}. 
We consider a plano-concave cavity with perfect rotation symmetry, use cylindrical coordinates defined by the symmetry $z$-axis, and denoted position by $(r,\theta,z)$. 
This cavity supports a set of matched Laguerre-Gaussian modes, with flat wavefronts at the flat mirror ($z=0$) and matched curved wavefronts at the concave spherical mirror ($z=L$, radius of curvature $R_m$).
The Rayleigh range $z_0 = {\scriptstyle \frac{1}{2}} kw_0^2 = \sqrt{L(R_m-L)}$ determines the beam waist $w_0$ at the planar mirror, where $E_{00}(r,\theta) \propto \exp(-r^2/w_0^2)$, the variation of the beam size upon propagation, via $w_z^2 =w_0^2 (1+z^2/z_0^2)$, the radius curvature of the wavefront $R(z) = R_z = z+z_0^2/z$, and the phase lag  $\chi(z)=\arctan(z/z_0)$ of the fundamental mode with respect to a plane wave. 
We split the intra-cavity standing wave in forward- and backward-propagating fields and write the slowly-varying component of the forward-propagating complex field of the Laguerre-Gaussian modes of this cavity as $E_{p,\ell} = \psi^+_{p,\ell} \exp{i(kz-\omega t)}$ with
\begin{equation}
\label{eq:psi-pl}
    \psi_{p,\ell}^+(r,\theta,z) = \frac{1}{\gamma_z} \Psi_{p,\ell}(\rho,\theta,\chi) \exp\left[ ik\frac{r^2}{2R_z} \right] \,. 
\end{equation}
The radial quantum number $p$ and OAM quantum number $\ell$ combine to the transverse order $N = 2p + |\ell|$.
The slowly-varying backward-propagating field, $\psi_{p,\ell}^-(r,\theta,z) = - (1/\gamma_z) \Psi_{p,\ell}(\rho,\theta,-\chi) \exp[-ikr^2/(2R_z)]$, is a mirror image of the forward-propagating field and hence will not be considered explicitly. 
%
In Eq.~(\ref{eq:psi-pl}), the normalized mode functions
\begin{equation}\label{Eq:normalized_mode_functions}
    \Psi_{p,\ell}(\rho,\theta,\chi) = f_{p,\ell}(\rho) e^{i\ell\theta} 
     \exp[ - i(N+1)\chi ] \,
\end{equation}
are the eigenmodes of the two-dimensional harmonic oscillator in quantum mechanics (QM), in terms of the  normalized transverse position $\rho = r/\gamma_z$, with $\gamma_z = w_z/\sqrt{2} = \gamma_0/\cos{\chi}$.
The amplitude functions 
\begin{equation}\label{Eq:amplitude_functions}
    f_{p,\ell}(\rho) = (-1)^p \sqrt{\frac{p!}{\pi (p+|\ell|)!}} \rho^{|\ell|} L^{|\ell|}_p\left(\rho^2\right)  \exp[ -\rho^2/2] \,,
\end{equation}
contain the generalized Laguerre polynomials $L^{|\ell|}_p$ and a sign $(-1)^p$ to make them equal to the transverse modes generated by harmonic oscillator ladder operators (see Appendix \ref{sec:Appendix-operators}). 
All modes are normalized via
\begin{subequations}
\begin{eqnarray}
    \langle \psi^+_{p,\ell} | \psi^+_{p,\ell} \rangle & \equiv & \int_0^\infty r dr \int_0^{2\pi} d\theta \,\, |\psi^+_{p,\ell}(r,\theta,z) |^2 = 1  \,, \\
    \langle \Psi_{p,\ell} | \Psi_{p,\ell} \rangle & \equiv & \int_0^\infty \rho d\rho \int_0^{2\pi} d\theta  \,\, |  \Psi_{p,\ell}(\rho,\theta,\chi) |^2 = 1 \,,
\end{eqnarray}
\end{subequations}
where $| \psi^+ \rangle = | \psi^+(z) \rangle$ is the ket notation for the mode $\psi^+(r,\theta,z) = \langle r, \theta | \psi^+(z) \rangle$.
These integrals do not depend on $z$, or $\chi$; the modes remain orthonormal because optical propagation is a unitary operation.

The resonance condition in a cavity is determined by the requirement that the mode reproduces itself after a roundtrip.
This requirement is satisfied when $R(L) = R_m$, which fixes $w_0$, and when the roundtrip phase $\varphi_{\rm round}$ is a multiple of $2\pi$, such that
\begin{equation}
\label{eq:resonance-phase}
    \varphi_{\rm round} = \varphi - \varphi_{\rm par} - \varphi_{\rm non} = 2\pi \, q \,,
\end{equation}
where $\varphi = 2 kL$ is the plane-wave roundtrip phase and $q$ is the longitudinal quantum number. 
The paraxial roundtrip phase lag $\varphi_{\rm par}$ equals $2 (N+1) \chi_0$.
The additional, typically small, roundtrip phase lag $\varphi_{\rm non}$ combines all non-paraxial and mirror-shape effects. 
Eq.~(\ref{eq:resonance-phase}) corresponds to resonant cavity lengths
\begin{equation}
\label{eq:resonance}
    L_j = \frac{\lambda}{2} \left[ q + (N+1) \frac{\chi_0}{\pi} + \Delta \tilde{\nu}_j \right] + 2 L_\varphi \,,
\end{equation}
where $\Delta \tilde{\nu}_j = \varphi_{\rm non}/(2\pi)$ and $L_\varphi$ is the phase penetration depth into the (Bragg) mirrors ($L_\varphi =0$ in the center of the stopband) \cite{Koks2021}.
Throughout this paper, we will quantify the magnitude of each non-paraxial and mirror-shape effect by its contribution to the dimensionless detuning $\Delta \tilde{\nu}_j$ of mode $j$.
The mode label $j=\{p,\ell,v\}$ combines the two transverse quantum numbers $p$ and $\ell$ with a third vector/polarization quantum number $v$; see Sec. \ref{sec:vector}.

\section{Roundtrip operator and perturbation theory}
\label{sec:Roundtrip}
\subsection{The roundtrip operator ${\cal M}$}
\label{sec:roundtrip-subsec}

The evolution of the optical field in a cavity can be described by the roundtrip operator $M$, which transforms the forward-propagating field $| \psi^+ \rangle$ into $M | \psi^+ \rangle$ after a roundtrip~\cite{Kleckner2010}. 
If we expand the field $| \psi^+ \rangle = \sum_j c_j | \psi_j^+ \rangle$ into a set of orthogonal basis states $|\psi^+_j \rangle$, then we can write $| \psi^+ \rangle$ as a vector and $M$ as a matrix. 
Finding the resonance conditions and eigenmodes of the cavity now boils down to finding the eigenvalues and eigenvectors of the roundtrip matrix. 

\begin{figure}
    \centering
    \includegraphics[width=0.6\linewidth]{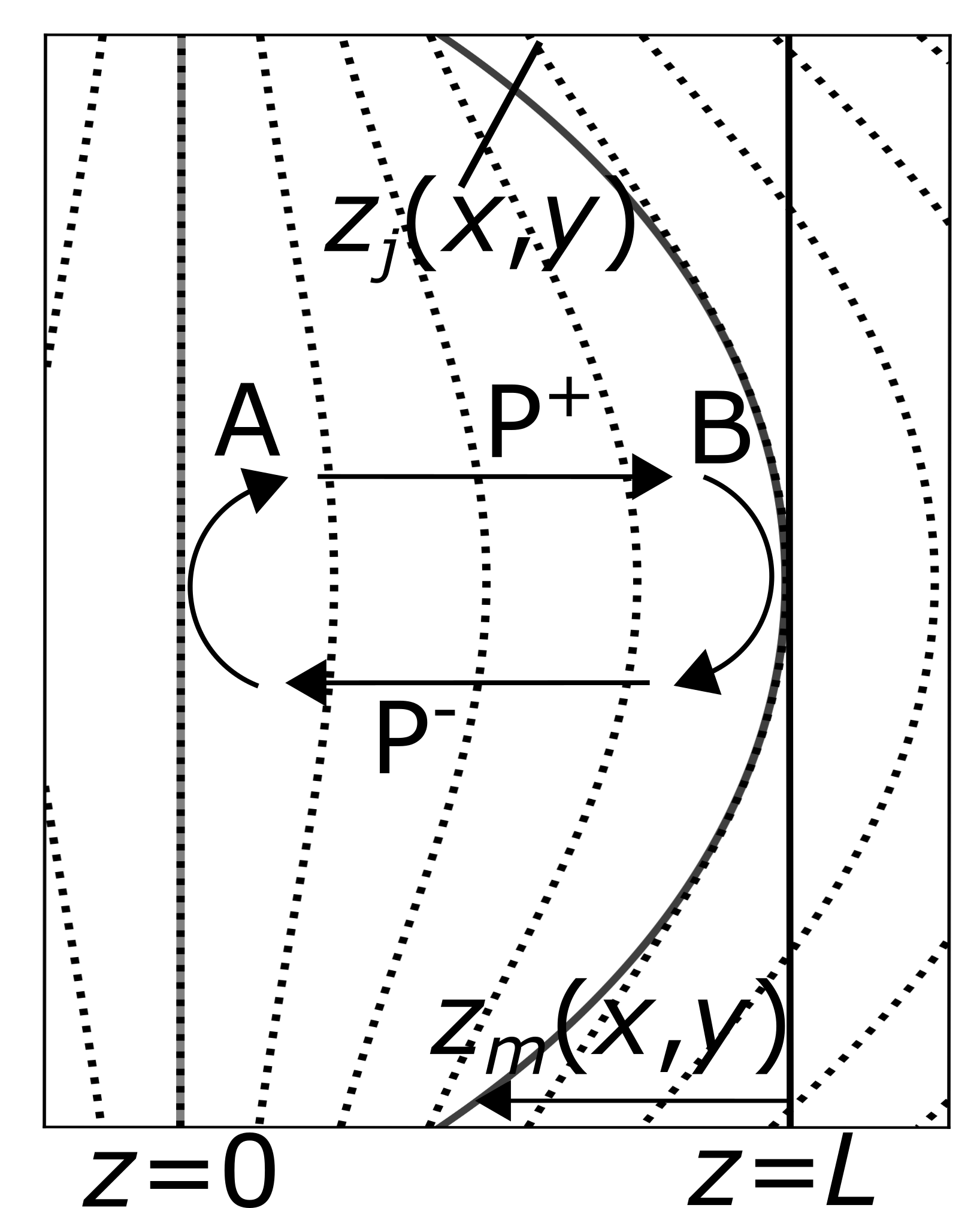}
    \caption{Geometry of a plano-concave cavity. Starting from the right, the optical roundtrip $M=P^+AP^-B$ includes reflection from the concave mirror $B$, propagation $P^-$ to the left, reflection from mirror $A$, and propagation $P^+$ to the right. The shape of the concave mirror is specified by its distance $z_m(x,y)>0$ from the plane $z=L$. The wavefront of the intracavity mode $j$ is flat at mirror $A$ and almost spherical, and described by $z_j(x,y)$, at mirror $B$.}
    \label{fig:geometry}
\end{figure}

The roundtrip operator in any two-mirror cavity can be written as $M=P^+AP^-B$, where $A$ and $B$ represent the reflections from the left and right mirrors and $P^+$ and $P^-$ represent the propagation from A to B and back (see Fig.~\ref{fig:geometry}).
In a plano-concave cavity with a large smooth planar mirror, $A$ is equal to unity and $M=PB$, where $P=P^+P^-$ describes the roundtrip propagation and $B$ describes the reflection from the concave mirror.

In a typical experiment, we illuminate the optical cavity with an input field $|\psi^+_{\rm in} \rangle$ through mirror A and we observe the output field $| \psi^+_{\rm out} \rangle$ through mirror $B$, as a function of the cavity length or the optical frequency.
The forward-propagating intra-cavity field $| \psi^+_{\rm cav} \rangle$ at mirror B can be calculated by summation over an infinite series of reflections and repeated operations of the roundtrip operator
\begin{equation}
\label{eq:FP}
    |\psi^+_{\rm cav} \rangle = (1 + M + M^2 + ...) t_1 P^+ |\psi^+_{\rm in} \rangle = \frac{t_1 P^+}{1-M} |\psi^+_{\rm in}\rangle \,,
\end{equation}
where $t_1$ and $t_2$ are the amplitude transmissions of mirrors A and B. 
Eq.~(\ref{eq:FP}) extends the single-mode treatment as found in many textbooks to a multi-transverse-mode treatment of the Fabry-Perot cavity, by 
treating $P^+$ and $M$ as operators instead of scalars, and $| \psi^+ \rangle$ as a field profile instead of a field amplitude. 
The output field is derived from the cavity field as $|\psi^+_{\rm out} \rangle = t_2 |\psi^+_{\rm cav} \rangle$.

The roundtrip operator $M$ determines the full dynamics of the intracavity field, as it contains the eigenfrequencies and damping rates of all cavity modes. 
To highlight this link, we write 
\begin{equation}
\label{eq:M}
    M = \exp(i\varphi-i{\cal H} -{\cal A}) \,,
\end{equation}
with Hermitian operators ${\cal H}$ and ${\cal A}$.
For the single-mode case, the scalar ${\cal H} = \varphi_j$ describes the roundtrip phase lag of mode $j$ with respect to the plane-wave roundtrip phase $\varphi = 2kL$.
The other scalar ${\cal A} = \gamma_j$ describes its roundtrip modal loss, where $\gamma_j = \gamma_r + \Delta \gamma_j$ combines the reflection loss $\gamma_r \approx 1 -r_1r_2$ (for amplitude reflectivities $r_{1,2}$ close to unity) 
with potential extra loss $\Delta \gamma_j$, like clipping loss from the finite-size mirrors.
For the multi-mode case, the dimensionless operator ${\cal H}$ describes the conservative dynamics, while the operator ${\cal A}$ describes the dissipative dynamics of all modes~\cite{Spreeuw1990,Suh2004}. 

The dynamics of the intra-cavity field is typically dominated by the modes that are close to resonance, i.e. modes for which $|| (1-M) |\psi^+_j \rangle || \approx 0$. 
We highlight this by expanding $(1-M) \approx i({\cal H}-\varphi)\, {\rm mod}(2\pi) + {\cal A}$, where the first term is taken modulo $2\pi$ to remove $\exp{i2\pi q}=1$. 
In the eigenbasis of the dynamics operator, defined by $M|\psi^+_j \rangle = \exp(i\varphi-i\varphi_j-\gamma_j)|\psi^+_j \rangle$, the output field of the multi-mode Fabry-Perot cavity can be written as
\begin{equation}
\label{eq:transmission}
       |\psi^+_{\rm out} \rangle = \sum_j \frac{t_1 t_2}{\gamma_j -i(\varphi-\varphi_j)} \, P^+ \, |\psi^+_j\rangle \langle \psi^+_j |\psi^+_{\rm in}\rangle \,,
\end{equation}
where $(\varphi-\varphi_j)$ is again taken modulo $2\pi$. 
Only modes close to resonance, with $(\varphi-\varphi_j) \approx 0 \, {\rm mod}(2\pi)$, contribute significantly to the transmission. 

The transmission spectrum described by Eq.~(\ref{eq:transmission}) is the counterpart of the dynamic equation
\begin{eqnarray}
\label{eq:time}
    T_{\rm round} \frac{d}{dt} | \psi^+ \rangle_X & = & | \psi^+ \rangle_{X+1} - | \psi^+ \rangle_X \\ & \approx & (M-1) \, | \psi^+ \rangle_X + t_1 P^+ | \psi^+_{\rm in} \rangle_X \nonumber \,,
\end{eqnarray}
which describes the evolution of the intra-cavity field $|\psi^+ \rangle$ from roundtrip $X$ to roundtrip $X+1$, in the roundtrip time $T_{\rm round}$.
This equation again shows that resonances in optical cavities behave like coupled harmonic oscillators, whose (complex) eigenvalues and eigenstates are determined by the resonance condition $({\cal H}-i{\cal A})| \psi_j \rangle = (\varphi_j - i \gamma_j) | \psi_j \rangle$. 

This equivalence was already discussed by 
Haus~\cite{Haus} and later extended by Fan et al.~\cite{Fan2003}. 
Suh, Wang and Fan \cite{Suh2004} expressed the dynamics of the field in any multi-mode cavity in its most general form as
\begin{equation}
\label{eq:Suh}
    \frac{d}{dt} \vec{a} = -(i\Omega+\Gamma) \vec{a} + \kappa^T | s_+ \rangle \,,
\end{equation}
where the vector $\vec{a}$ combines the (complex) amplitudes of all relevant cavity modes, the $\Omega$ and $\Gamma$ are dynamic matrices, and where the matrix $\kappa^T$ couples the (multi-channel) input field $| s_+ \rangle$ to the cavity modes. 
Our Eq.~(\ref{eq:time}) resembles Eq.~(\ref{eq:Suh}), but also differs in three ways: 
first of all, Eq.~(\ref{eq:time}) expresses the intra-cavity field $| \psi^+ \rangle$ and the input field $| \psi^+_{\rm in} \rangle$ as two 2D field profiles, whereas $\vec{a}$ in Eq.~(\ref{eq:Suh}) is a 3D intra-cavity field while $| s+ \rangle$ is a 2D field profile. 
This difference introduces a square root of time in the dimension of $\kappa^T$.  
Second, our Eq.~(\ref{eq:time}) describes the dynamics of the slowly-varying field $\psi^+(t)$, whereas Eq.~(\ref{eq:Suh}) describes the dynamics of the full field $E(t)$; 
this difference results in a factor $\exp(i\varphi)$ in $M$.
Finally, and most importantly, our  Eq.~(\ref{eq:time}) links the dynamics of the intra-cavity field to the roundtrip operator $M$; 
we will use this link to determine the intra-cavity modes and their properties. 

\subsection{Perturbation theory applied to ${\cal H}$}
\label{sec:perturbation-theory}

From now on, we will neglect losses, assuming ${\cal A} \approx 0$. 
We will remove the + subscript and interpret $| \psi \rangle$ as the forward-propagating field at the curved mirror, i.e. $| \psi \rangle = | \psi^+(z=L) \rangle$. 
This is allowed when resonances are sharp and when we are not interested in their spectral width.
We split the dynamic operator as ${\cal H} = {\cal H}_{\rm par} + {\cal H}_{\rm fine}$ and associate the resonance with a hard zero in the resonance condition
\begin{eqnarray}
\label{eq:resonance2}
    (M-1)|\psi \rangle & = & (\exp{i(\varphi-{\cal H})} - 1) | \psi \rangle \,\, \nonumber \\
    & \approx & i (\varphi - q2\pi - {\cal H}_{\rm par} - {\cal H}_{\rm fine}) | \psi \rangle = 0 \,,
\end{eqnarray}
where $q$ is the longitudinal mode number. 
The operator ${\cal H}_{\rm par}$ describes the paraxial evolution in a cavity with a spherical concave mirror.
The operator ${\cal H}_{\rm fine}$ describes the, typically small, modifications due to non-paraxial propagation and reflection from a non-spherical mirror.

The dynamic matrix ${\cal H}_{\rm par}$ is diagonal in the basis of the paraxial eigenmodes presented in Sec. \ref{sec:paraxial}.
The on-diagonal elements are equal to the roundtrip phase lag $\varphi_j = 2(N+1) \chi_0$ and are thus identical for modes with the same transverse order $N$.
Each $N$-group contains $2(N+1)$ modes, divided over $(N+1)$ spatial profiles, labelled by $(p,\ell)$ for the scalar LG-modes, times two polarizations.

The fine structure operator ${\cal H}_{\rm fine}$ can lift the frequency degeneracy within each $N$-group and reshape the eigenmodes.
We calculate these effects by applying perturbation theory to Eq.~(\ref{eq:resonance2}). 
In principle, this perturbing operator can couple and mix all paraxial modes.
In practice, it mainly mixes modes of the same order $N$. 
The fine structure within each $N$-group is thus described by first-order frequency-degenerate perturbation theory and by the transverse-mode matrix 
\begin{equation}
\label{eq:nu-matrix1}
    \varphi_{{\rm fine},j'j} = 2\pi \Delta \tilde{\nu}_{j'j} = \langle \psi_{j'} | {\cal H}_{\rm fine} | \psi_j \rangle \ll 1 \,.  
\end{equation}
The eigenvalues of the matrix $\Delta \tilde{\nu}$ yield the spectral shifts, i.e. the fine structure.
The eigenvectors of $\Delta \tilde{\nu}$ yield the new eigenmodes of the cavity.

The coupling between modes from different transverse orders is far less effective and described by second-order frequency non-degenerate perturbation theory \cite{Sakurai}.
This yields coupling rates of the form $\Delta \tilde{\nu}_{N \neq N'} \approx \pi (\Delta \tilde{\nu}_{j',j})^2/(N-N')\chi_0$, where $(N-N')\chi_0/\pi$ is the frequency difference between the paraxial modes in the $N$ and $N'$ groups. 
Non-resonant coupling is strongly suppressed by the denominator, $(N-N')\chi_0 \gg \pi \Delta \tilde{\nu}_{j',j}$, and can typically be neglected. 
From a physics point of view, the extra field $-i{\cal H}_{\rm fine} |\psi \rangle$ in Eq. (\ref{eq:resonance2}) only remains trapped in the cavity when it fits resonantly in the cavity and light scattered to other modes does not build up resonantly, is quickly lost, and can thus be neglected.
The only exception to this rule is the situation where modes of different orders are accidentally almost frequency-degenerate~\cite{Koks2022a}; we will not consider this case any further.  

\begin{table*}[tbh]
    \centering
    \begin{tabular}{|c|l|c|c|c|c|c|}
    \hline 
     {\bf Category} & {\bf Contribution} & {\bf {\cal H}} & {\bf Form} & {\bf Preferred basis} & {\bf Magnitude} $\varphi_j = 2\pi \Delta \tilde{\nu}_j$ & {\bf Discussed in Sec.} \\ \hline \hline
     Paraxial & Paraxial  & par & $r^2$ and $k_\perp^2$ & no preference & $2(N+1)\arcsin{\sqrt{L/R_m}}$ & \ref{sec:paraxial}   \\ \hline \hline
     & Non-paraxial scalar &  & &  &  &    \\
     (known) & - Propagation/Helmholtz & prop & $k_\perp^4$ & scalar LG & magnitude prop+wave: & \ref{sec:prop} \\ 
     non- & - Wavefront/Spherical & wave & $r^4$ & scalar LG & $[g(p,\ell)+4]/(4 kR_m)$ & \ref{sec:wave}\\ 
     paraxial & Mirror aspheric & asphere & $r^4$ & scalar LG & $- \tilde{p} f(p,\ell) L/[4 kR_m(R_m-L)]$ &  \ref{sec:asphere} \\ 
      & Non-paraxial vector & vec & $\vec{r} \otimes \vec{k}_\perp$ & vector LG & $[-1 - \ell \cdot s]/(kR_m)$  & \ref{sec:vector-sub} \\ 
      & = spin-orbit coupling &  &  &  &  & \\ \hline \hline
     (new) & Mirror Bragg correction & bragg & $\vec{k}_\perp \otimes \vec{k}_\perp$ & vector LG & $\pm C_0(N+1)/(kz_0)$ & \ref{sec:Bragg} \\ 
     vector & &  &  &  & hyperfine for $1A\pm$ modes &    \\ 
      &  &  & $k_\perp^2 \, (\vec{k}_\perp \otimes \vec{k}_\perp)$  &  & $\propto C_2$ (not analyzed in detail) & \\ \hline \hline
     (new) & Mirror astigmatism & astigm & $x^2-y^2$ & scalar HG & $ \eta_{\rm astigm} \sqrt{L/(R_m-L)}$ & \ref{sec:astigm} \\ 
     scalar &  &  &  &  & (off diagonal, $\Delta \ell = \pm 2$) & \\ \hline \hline
     (new) & Mirror astigmatic vector & v+a & & x- and y-pol. & $ \pm \eta_{\rm astigm} /(kR_m)$ & \ref{sec:shape}  \\     
     vector & = anisotropic spin-orbit &  &  &  & hyperfine for $\ell=0$ modes &  \\ \hline 
    \end{tabular}
    \caption{Overview of various contributions to the transverse mode spectrum, with their abbreviated name, operator form, preferred basis, relative strength (= non-paraxial phase lag) $\varphi_j$, and associated section. The paraxial contribution has no preferred basis; the rotation-symmetric corrections prefer the LG-basis; mirror astigmatism prefers the HG-basis.}
    \label{tab:table}
\end{table*}

\subsection{Symmetry aspects \& scalar versus vector modes}
\label{sec:symmetry} 

The eigenmodes of the fine structure operator ${\cal H}_{\rm fine}$ can often already be determined from the symmetry of the system. 
For cavities with rotation symmetry, each scalar eigenmode has a fixed OAM, with quantum number $\ell$, and each vector mode has a circular polarization $\sigma_\pm$, with spin quantum number $s=\pm 1$.
For cavities with additional mirror symmetry, as is common, the $\ell$ and $-\ell$ eigenmodes should be frequency degenerate.
From now on, we will use this argument repeatedly and take $\ell \geq 0$ throughout the main text.
Hence, the $\ell \neq 0$ modes are expected to form frequency-degenerate groups of four polarized modes, while the $\ell = 0$ modes are expected to form polarization pairs.
Below, we will show that spin-orbit coupling breaks each $\ell \neq 0$ group of four modes into two pairs of vector modes.
We will also show that the final pairwise degeneracy is more difficult to break.
In analogy with the atomic fine structure, we propose to call the final pairwise break-up the hyperfine component of the fine structure. 
But first, we will discuss general symmetry aspects of ${\cal H}_{\rm fine}$, based on the distinction between scalar versus vector effects and between rotation-symmetric versus astigmatic cavities. 

The calculation of the scalar corrections is based on the idea that the extra field $-i{\cal H}_{\rm fine} |\psi \rangle$ results from the mismatch between the shape of the concave mirror and the wavefront of the mode.
The reflected field is then multiplied by a factor $\exp(2ik\Delta z) \approx 1+2ik\Delta z$ and ${\cal H}_{\rm fine} = - 2k\Delta z$, or actually ${\cal H}_{\rm fine} = + 2k\Delta z$ with our sign definition, where $\Delta z(x,y) = z_{\rm mirror} - z_{\rm wave}$ and positive $z_{\rm mirror}$ and $z_{\rm wave}$ point towards the plane mirror (see Fig.~\ref{fig:geometry}).  
Substitution of ${\cal H}_{\rm fine} = 2k\Delta z$ into Eq.~(\ref{eq:nu-matrix1}) results in a dynamics matrix of the form
\begin{eqnarray}
\label{eq:modal-integration2}
    \Delta \tilde{\nu}_{j'j} & = & \frac{2}{\lambda} \langle \psi_{j'} | \Delta z | \psi_j \rangle \\ 
    & = & \frac{2}{\lambda} \iint   \mbox{d}x \mbox{d}y \, \Delta z(x,y)\, \psi^*_{j'}(x,y) \psi_j(x,y) \nonumber \,,
\end{eqnarray}
where $\Delta z = z_{\rm mirror} - (z_{j'} + z_{j})/2$.
The complex conjugation removes the curvature and Gouy phase from $\psi(x,y)$. 

For a rotation-symmetric cavity, this matrix is diagonal in the eigenbasis of the scalar LG-modes and the fine structure is given by 
\begin{equation}
\label{eq:modal-integration}
    \Delta \tilde{\nu}_j = \frac{2}{\lambda} \langle \psi_j | \Delta z | \psi_j \rangle = \frac{2}{\lambda} 
    \iint   \mbox{d}x \mbox{d}y \, \Delta z(x,y)\ |\psi_j(x,y)|^2  \,,
\end{equation}
where $\Delta z = z_{\rm mirror} - z_j$ is the mismatch between the shape of the mirror $z_{\rm mirror}$ and the wavefront $z_j$ of the paraxial LG-mode $j$. 
Equations~(\ref{eq:modal-integration2}) and~(\ref{eq:modal-integration}) can also be derived by applying the theory of Kleckner et al.~\cite{Kleckner2010} to a plano-concave cavity with $\Delta z \ll \lambda/2$ (see Appendix of Ref.~\cite{Koks2022a}), but we think the derivation presented above is easier.

Let's consider two simple examples of Eq.~(\ref{eq:modal-integration}).
As the first example, we consider a uniform displacement of the curved mirror towards the plane mirror over $\Delta z = \alpha >0$.
Substitution in Eq.~(\ref{eq:modal-integration}) now yields $\Delta \tilde{\nu}_j = 2\alpha/\lambda$ for all transverse modes and substitution in Eq.~(\ref{eq:resonance2}) shows that the resonant cavity length increases by $\Delta L_j = \alpha$ for all modes, as expected.
As the second example, we analyze the effect of a small increase in the mirror curvature, described by $\Delta z = \beta r^2$ with $\beta >0$, making $\Delta R_m = -2\beta R_m^2 <0$, and $|\Delta R_m| \ll R_m$.
Substitution into Eq.~(\ref{eq:modal-integration}) now yields $\Delta \tilde{\nu}_j = (2/\lambda) \times \beta (N+1)(w_1^2/2)$. 
Substitution in Eq.~(\ref{eq:resonance}) again yields the associated change in the cavity length, which now equals $\Delta L_j = \beta (N+1)(w_1^2/2) > 0$. 
This result can be fully attributed to the change in the Gouy phase $(N+1)\chi_0$ due to the increased $R_m$ at fixed $L$. 
Using $\chi_0 = \arcsin{\sqrt{L/R_m}} \Rightarrow \Delta \chi_0 = -\sqrt{L/(R_m-L)} \Delta R_m/(2R_m)$ and the expression for $w_1^2=w_z^2$ at $z=L$ presented in Sec.~\ref{sec:paraxial}, we again find full agreement.

The calculation of vector corrections requires an extension from scalar modes $| \psi_j \rangle$ to vector modes $| \vec{\psi}_j \rangle$ and from scalar operators ${\cal H}_{\rm fine}$ to $2 \times 2$ tensor operators.
We will discuss two different vector corrections in Secs.~\ref{sec:vector} and~\ref{sec:Bragg} and show that: (i) the spin-orbit coupling is relatively strong and present for all $\ell \neq 0$ modes and (ii) the Bragg correction is typically weak and mainly observable for some $\ell=1$ modes. 

In Sec.~\ref{sec:astigm}, we will analyze astigmatic cavities, without rotation symmetry.  
We will show how astigmatism modifies the eigenvalues and eigenmodes, by coupling modes with different $\ell$, and how it retains the two-fold degeneracy of the $\ell \neq 0$ modes while creating a small (second-order) frequency splitting of the $\ell = 0$ pair. 

\subsection{Contributions to the fine structure (Table~I)}
\label{se:contributions}
Before discussing in detail the announced vector corrections and effects of astigmatism etcetera, we present a brief overview in a table of the various contributions to the transverse mode spectrum that are covered in this paper. 
The optical fine structure has many contributions, which are linked to different physical processes and described by different contributions to the fine structure operator ${\cal H}_{\rm fine}$. 
Table~I lists the most relevant contributions and compares their properties and approximate strengths.
The contributions are divided in four categories/blocks: 
The first block describes the effect of the paraxial operator ${\cal H}_{\rm par}$.
The second block describes effects that occur in cavities with rotational symmetry. 
These effects are divided in two non-paraxial scalar effects ${\cal H}_{\rm scalar} = {\cal H}_{\rm prop} + {\cal H}_{\rm wave}$, an aspherical mirror effect ${\cal H}_{\rm asphere}$, and a non-paraxial vector effect ${\cal H}_{\rm vec}$.
The third block describes a vector effect ${\cal H}_{\rm Bragg}$ that occurs in rotation-symmetric cavities with Bragg mirrors. 
The fourth block describes the effects of astigmatic mirrors, divided in the dominant effect ${\cal H}_{\rm astigm}$ and a second-order effect ${\cal H}_{\rm v+a}$. 
The effects in Block~Two have been discussed in the literature, albeit often as individual effects and in different notations; the analysis of the effects in Blocks Three and Four is new.
All mentioned effects are typically small and hence simply add up, albeit as matrices if they prefer different bases. 

The first three columns in Table~I show the names of the various effects and their history. 
The fourth column shows the functional form of the associated operator ${\cal H}_{\rm fine}$.
The fifth column shows the preferred eigenmodes, which are Laguerre-Gauss (LG) scalar or vector modes for rotation-symmetric cavities and Hermite-Gauss (HG) modes for astigmatic cavities. 
The sixth column quantifies the relative strengths of the expected effects.
The final column refers to the sections in which each effect is discussed. 
Sections~\ref{sec:scalar} and ~\ref{sec:vector} describe three scalar corrections and the spin-orbit vector correction for rotation-symmetric cavities. 
Together, they present the ``known" non-paraxial rotation-symmetric corrections discussed in part II of this paper. 
Section~\ref{sec:Bragg} describes the vector Bragg correction for rotation-symmetric cavities with Bragg mirrors, while Sec.~\ref{sec:mirror} analyzes two scalar corrections in astigmatic cavities. 
Together, they present the ``new" non-paraxial effects that form the basis of part III of this paper. 

\begin{figure}[tbh]
    \centering
    \includegraphics[width=\linewidth]{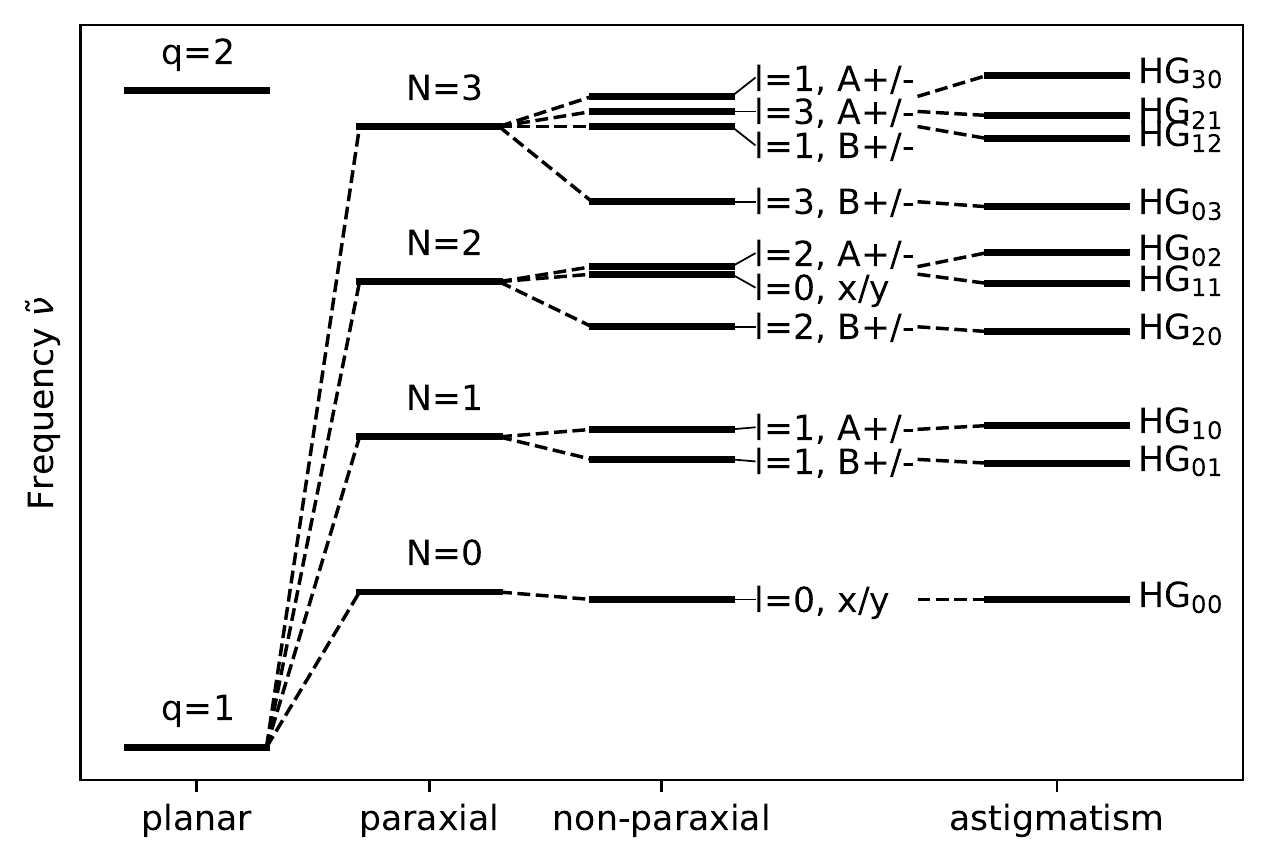}
    \caption{Sketch of the expected resonance frequencies of an optical cavity (vertical axis) under the influence of various perturbations (horizontal axis). Left column: modes in a planar cavity. Second column: paraxial modes in a plano-concave cavity. Third column: fine structure in a rotational-symmetric cavity. Right column: modified fine structure in an astigmatic cavity.}
    \label{fig:finestructure}
\end{figure}

Figure \ref{fig:finestructure} shows a sketch of the changes expected for the different contributions.
The first two columns show how the paraxial resonances in a planar and a plano-concave cavity cluster in groups with the same quantum numbers $q$ and $N$. 
The third column shows how non-paraxial scalar and vector corrections split these clusters into pairs of modes with additional quantum numbers $\ell$ and $v$. 
The right column shows how deviations from rotation symmetry, due to astigmatic mirrors, will modify both the fine structure and the character of the eigenmodes, from LG-modes to HG-modes. 
The third and final column depict the fine structure mentioned in the title. 
A detail that is not visible in the figure is that each line consists of two, typically frequency degenerate, modes that sometimes exhibit a tiny hyper-fine splitting. 
Below we will show that these hyperfine splitting originate from mixing of optical polarizations and that the $\ell =0$ and $\ell=1,A$ modes are most susceptible to hyperfine splittings.

We end this section by addressing the completeness of our list of non-paraxial effects. 
We first note that operators that are 3$^{\rm rd}$-order in $\vec{r}$ and $\vec{k}_\perp$ are irrelevant for cavities with inversion symmetry, as their effects average to zero.
But why did we single out the listed effects as the dominant ones, and why did we choose not to include other $2^{\rm nd}$ and $4^{\rm th}$-order contributions to ${\cal H}_{\rm fine}$ that are also allowed by symmetry?
Our reasoning is as follows:
A potential contribution of the $4^{\rm th}$-order operator $r^2 k_\perp^2$, whatever its physical mechanism, is probably much weaker than that of the related 2$^{\rm nd}$-order operator $\vec{r} \otimes \vec{k}_\perp$ and has hence been neglected. 
The same argument applies to the operator $x^4-y^4$, which describes the non-paraxial contribution to the astigmatism.
Furthermore, we find it hard to envision a physical mechanism for the $\vec{r} \otimes \vec{r}$ operator.
And individual $k_x^2-k_y^2$ and $k_xk_y$ operators are probably only relevant in birefringent cavities.
Hence, we think our list is complete for most practical purposes. 

\section{Non-paraxial scalar corrections}
\label{sec:scalar}

\subsection{Helmholtz correction ${\cal H}_{\rm prop}$}
\label{sec:prop}

The Helmholtz correction ${\cal H}_{\rm prop}$ originates from a non-paraxial contribution to the propagation. 
Non-paraxial propagation has been studied extensively;  Ref.~\cite{Sheppard1999} gives a brief historic overview. 

We base the first part of our analysis on the work of Lax~\cite{Lax1975}, who starts by noting that the intra-cavity optical field $\vec{E}(\vec{r})$ must satisfy Maxwell's equations, which for a monochromatic field reduce to the vector Helmholtz equation $(\nabla^2 + k^2) \vec{E}(\vec{r}) = \vec{0}$ and the divergence condition $\vec{\nabla} \cdot \vec{E}(\vec{r})=0$.
He then introduces the slowly-varying forward-propagating field, which we write as $\vec{E} = (\vec{\psi}_\perp + \psi_z \vec{e}_z) \exp{i(kz-\omega t)}$, where $\vec{\psi}_\perp$ combines the transverse components $\psi_x$ and $\psi_y$ and $\psi_z$ is the axial/longitudinal component. 
And he expresses this field as a Taylor expansion of the paraxial field and a series of non-paraxial corrections, with amplitudes that decay as a power series in the expansion parameter $f =1/(kw_0) = \Theta_0/2\ll 1 $, where $w_0$ is the waist and $\Theta_0$ is the opening angle of the fundamental paraxial mode.
The resulting equations are 
\begin{eqnarray}
\label{eq:paraxial}
    \left( \Delta_\perp + 2ik \frac{d}{dz}\right) \vec{\psi}_\perp^{(0)}(\vec{r}) & = & 0 \,, \\
\label{eq:longitudinal}
    \psi_z^{(1)}(\vec{r}) & \approx & \frac{i}{k} \left( \frac{d \psi_x^{(0)}}{dx} + \frac{d \psi_y^{(0)}}{dy}  \right) \,, \\
\label{eq:Helmholtz-modified}
    \left( \Delta_\perp + 2ik \frac{d}{dz}\right) \vec{\psi}_\perp^{(2)}(\vec{r}) & = & - \frac{d^2}{dz^2} \vec{\psi}_\perp^{(0)} \approx \frac{\Delta_\perp^2}{4k^2} \, \vec{\psi}_\perp^{(0)} \,,  
\end{eqnarray}
where $\vec{\nabla}_\perp = \partial_x \vec{e}_x + \partial_y \vec{e}_y$ is the transverse nabla operator and $\Delta_\perp = \vec{\nabla}_\perp^2 = \partial_x^2 + \partial_y^2$.
These three equations describe, respectively, the evolution of the paraxial field $\vec{\psi}^{(0)}_\perp$, the axial field $\psi_z^{(1)}$, and the extra (non-paraxial) transverse field $\vec{\psi}^{(2)}_\perp$.

The scaling/expansion argument of Lax~\cite{Lax1975} is as follows:
For Gaussian beams with a waist $w_0$, the `transverse' derivatives $d/dx$ and $d/dy$ in Eq.~(\ref{eq:longitudinal}) will generate factors of the order $1/w_0$.
Hence, the axial field $\psi_z^{(1)}$ is approximately a factor $f=1/(kw_0)$ smaller than the transverse field $\vec{\psi}_\perp^{(0)}$.
And the extra `axial' derivative $d/dz$ in Eq.~(\ref{eq:Helmholtz-modified}) will generate a factor of the order $1/z_0$. 
Hence, the non-paraxial Helmholtz correction to the transverse field, $\vec{\psi}^{(2)}_\perp$, will be about a factor $f^2/4=1/(2kw_0)^2=1/(8kz_0)$ smaller than the original paraxial transverse field. 
By ending the Taylor expansion after the $f^2$-term, we neglect an even smaller $f^3$ correction on $\psi_z$ and an $f^4$ correction on $\vec{\psi}_\perp$. 

To calculate the non-paraxial scalar correction, we use a procedure introduced by Erickson~\cite{Erickson}.
We consider the propagation of a paraxial LG-mode $j$ from the flat to the concave mirror and expand the propagating field in the basis of LG-modes as~\cite{Erickson}
\begin{equation}
    | \psi(z) \rangle = | \psi_j \rangle + \sum_{j'} c_{j'}(z) | \psi_{j'}(z) \rangle \,,
\end{equation}
with $c_{j'}(0)=0$ and $c_{j'}(z) \ll 1$.
Substitution in Eq.~(\ref{eq:Helmholtz-modified}) and projection on $| \psi_{j'}(z) \rangle$ shows that non-paraxial propagation modifies the paraxial modes by 
\begin{equation}
\label{eq:c-z}
    \frac{d c_{j'}(z)}{dz} = \frac{-i}{8k^3} \, \langle \psi_{j'} | \Delta_\perp^2 | \psi_j(z) \rangle \,.
\end{equation}
Erickson~\cite{Erickson} has shown that the action of the $\Delta_\perp^2$ operator, or the related $\frac{d^2}{dz^2}$ operator, on LG-modes changes the radial index $p$ to $p'=p-2,p-1,p,p+1,p+2$, while leaving $\ell$ unchanged on account of the rotation symmetry.
But we are only interested in the $p'=p$ term, as the coupling to modes with different order $N$ is non-resonant.

Equation~(\ref{eq:c-z}) becomes intuitive when we write $\langle \psi_j | \Delta_\perp^2 | \psi_j \rangle = \langle k_\perp^4 \rangle$.
We then find that the correction $dc_j(z)/dz$ to the propagation $ik_z$ originates from the third term in the Taylor expansion $k_z = \sqrt{k^2-k_\perp^2} = k - k_\perp^2/(2k) - k_\perp^4/(8k^3)$.
As LG-modes retain their functional form under Fourier transformation, one easily finds $\langle \psi_{p,\ell} | \Delta^2_\perp | \psi_{p,\ell} \rangle = \langle k_\perp^4 \rangle = 4 f(p,\ell)/w_0^4$, where $f(p,\ell)$ is defined in Eq.~(\ref{eq:fpl}).
This result, and many others, can also be derived with the operator algebra described in Appendix~\ref{sec:Appendix-operators}.
Substitution in Eq.~(\ref{eq:c-z}) yields the relative frequency shift of the cavity resonances $\Delta \nu_j/\nu = - \Delta k_j/k = f(p,\ell)/(2k^4w_0^4) > 0$.
Conversion to a normalized frequency $\Delta\tilde{\nu}_j = (2L/\lambda) \Delta \nu/\nu$, yields the Helmholtz contribution to the fine structure
\begin{equation}
\label{eq:Erickson}
    \Delta \tilde{\nu}_{{\rm prop},j} = \frac{1}{8\pi k} \frac{L}{z_0^2} f(p,\ell) = \frac{1}{8\pi k R_m} \frac{w_1^2}{w_0^2} f(p,\ell) \,,
\end{equation}
in terms of the polynomial~\cite{Yu1984}
\begin{eqnarray}
\label{eq:fpl}
    f(p,\ell) & = & 6p^2+6p\ell+\ell^2+6p+3\ell+2 \nonumber \\
    & = & \frac{3}{2}(N+1)^2 - \frac{1}{2} (\ell^2-1) \,,
\end{eqnarray}
with $\ell \geq 0$.
This result is consistent with earlier results of Erickson \cite{Erickson,Erickson1977}, Yu and Luk~\cite{Yu1983,Yu1984}, and Luk~\cite{Luk1986}.
%
%
%

\subsection{Wave-front correction ${\cal H}_{\rm wave}$} 
\label{sec:wave}

The wave-front correction ${\cal H}_{\rm wave}$ originates from the difference between the optical wavefront and a reference surface.
On first sight, one might think that the paraxial wavefront should be parabolic, because the optical phase $\phi(r,z) = kz - (N+1)\chi + kr^2/(2R)$ in Eq.~(\ref{eq:psi-pl}) increases quadratically with $r$ in any $z$-plane. 
But wavefronts are defined by surfaces of fixed phase, and both $\chi$ and $R$ are functions of $z$. 
To find the true wavefronts, we use the pragmatic approach of Yu and Luk \cite{Yu1984} by expanding $\chi(z)$ and $R(z)$ around the paraboloidal reference surface $z = L - r^2/(2R_m)$ that one would naively expect, to find \cite{Yu1984}
\begin{equation}
\label{eq:z-wave}
    z_{\rm wave}(r) = \frac{r^2}{2R_m} + \frac{2(N+1)}{k^2w_1^2} \frac{r^2}{2R_m} - \frac{r^4}{4R_m^2L} \left( 1 - \frac{2L}{R_m} \right) \,,
\end{equation}
where $z_{\rm wave}(z) > 0$ for displacements from the $z=L$ plane towards the flat mirror; see Fig. \ref{fig:geometry}.
Incidentally, an alternative Taylor expansion of $\chi(z)$ and $R(z)$ around a spherical, instead of a paraboloidal, surface would yield a similar result, as the extra terms are relatively small for $L \ll R_m$.
The first term in Eq.~(\ref{eq:z-wave}) yields the curvature that we started from.
The second term shows that the central parts of the wavefronts are actually more curved, by a relative amount $2(N+1)/(kw_1)^2$. 
The final $r^4$-term makes the outer regions of the paraxial wavefronts ``flatter than paraboloidal" for the typical case $L < R_m/2$, and even further away from spherical. 

We compare the paraxial wavefronts with the surface of a spherical mirror
\begin{equation}
\label{eq:z-mirror}
    z_{\rm mirror}(r) = R_m - \sqrt{R_m^2-r^2} \approx \frac{r^2}{2R_m} + \frac{r^4}{8R^3_m} \,. 
\end{equation}
The mismatch $\Delta z = z_{\rm mirror} - z_{\rm wave}$ results in a shift of the resonance by an amount $\Delta \tilde{\nu}_{\rm wave} = (2/\lambda) \langle \psi_j | \Delta z | \psi_j \rangle$. 
We substitute Eqs.~(\ref{eq:z-wave}) and (\ref{eq:z-mirror}) into Eq.~(\ref{eq:modal-integration}), to find 
\begin{eqnarray}
\label{eq:scalar2}
    \Delta \tilde{\nu}_{{\rm wave},j} & = & -  \frac{(N+1)^2}{2\pi kR_m} + \frac{f(p,\ell)}{8\pi kR_m} \left( 3 - \frac{w_1^2}{w_0^2} \right) \nonumber \\
    & = & \frac{1}{8\pi k R_m} \left[ g(p,\ell) + 4 - \frac{w_1^2}{w_0^2} f(p,\ell) \right] \,,
\end{eqnarray}
where the combination~\cite{Yu1984}
\begin{eqnarray}
\label{eq:g}
    g(p,\ell) + 4 & = & 2p^2+2p \ell - \ell^2 + 2p + \ell + 2 
    \nonumber \\
    & = & \frac{1}{2}(N+1)^2 - \frac{3}{2} (\ell^2-1) \,.
\end{eqnarray}
To obtain this result we used $\langle r^2 \rangle = (N+1)w_1^2/2$, $\langle r^4 \rangle = f(p,\ell) w_1^4/4$, and $w_1^2/w_0^2= R_m/(R_m-L)$, and wrote some combinations of $L$ and $R_m$ in terms of the beam waists $w_0$ and $w_1$ at the two mirrors.
Our Eq.~(\ref{eq:scalar2}) is identical to Eq.~(27) in Ref.~\cite{Luk1986} and consistent with Eq.~(19) in Ref.~\cite{Yu1984}.

The combination of Eqs.~(\ref{eq:Erickson}) and (\ref{eq:scalar2}) finally yields the total non-paraxial scalar correction for a plano-concave cavity with a spherical mirror
\begin{equation}
\label{eq:scalar-total}
    \Delta \tilde{\nu}_{{\rm scalar},j} = \Delta \tilde{\nu}_{{\rm prop},j} + \Delta \tilde{\nu}_{{\rm wave},j} = \frac{1}{8\pi k R_m} \left[ g(p,\ell) + 4 \right] \,.
\end{equation}
That the sum of the two scalar corrections does not depend on the cavity length suggests an underlying physical reason, but we have not found it yet. 

\subsection{Aspherical correction ${\cal H}_{\rm asphere}$}
\label{sec:asphere}

In the previous section we calculated the wave-front correction by comparing the shape of the wavefront with a spherical mirror. 
We will now calculate the effect of a deviation from this spherical mirror shape by an amount 
\begin{eqnarray}
\label{eq:asphere-surface}
    z_{\rm mirror}(x,y) - z_{\rm sphere}(x,y) = - \tilde{p} \frac{r^4}{8R_m^3}\,,
\end{eqnarray}
where $\tilde{p}=0$ for a sphere (our reference) and $\tilde{p}=1$ for a paraboloid. 
This aspherical correction modifies the third term in the Taylor expansion $z_m(r) = a + b r^2 + c r^4 + ...$ of a rotation-symmetric mirror. 
Substitution of Eq.~(\ref{eq:asphere-surface}) in the generic Eq.~(\ref{eq:modal-integration}) yields the aspherical correction
\begin{equation}
\label{eq:asphere}
    \Delta \tilde{\nu}_{\rm asphere, j} = -  \frac{f(p,\ell)}{8\pi kR_m} \, \tilde{p} \, \frac{L}{R_m-L} \,.
\end{equation}
This result is consistent with the result of Zeppenfeld and Pinkse~\cite{Zeppenfeld2010}, who chose the paraboloidal mirror as their reference shape instead, and used a different notation;
see Appendix~\ref{sec:Appendix-Zeppenfeld}.

A comparison between Eq.~(\ref{eq:asphere}) with Eq.~(\ref{eq:scalar-total}) shows that the aspherical correction contains an extra factor $L/(R_m-L)$. 
The aspherical correction is thus relatively small for short cavities, basically because the modes in these cavities are relatively compact and hence less sensitive to mirror deformations. 

\section{Vector correction \& L-S coupling}
\label{sec:vector}
\subsection{Vector-LG modes}
\label{sec:vector-LG}
The analysis presented above used the scalar LG-modes $| \psi_{p,\ell} \rangle$ as basis set. 
This section extends the analysis to vector fields, by including the optical polarization.
It starts by introducing the vector LG-modes $| \vec{\psi}_{p,\ell,v} \rangle$, with their additional vector quantum number $v$.

In a cavity with mirror and rotation symmetry, the paraxial scalar modes $\psi_{p,\ell}$ and $\psi_{p,-\ell}$, with $\ell \geq 0$ are frequency degenerate. 
For vector fields, one also expects $x$- and $y$-polarized vector versions of these modes, such that the $\ell =0$ mode is two-fold degenerate and the $\ell \neq 0$ modes are four-fold degenerate. 
But in reality, the four $\ell \neq 0$ modes couple and split into two frequency-degenerate pairs which differ in the orientation of the photon spin $s$, or circular polarization $\sigma_\pm$, with respect to the orbital angular momentum $\ell$ due to a form of L-S coupling (see Sec.~\ref{sec:vector-sub}).

We will use the notation of Yu and Luk~\cite{Yu1983} and label the resulting vector LG-modes as: (i) series A modes with total angular momentum $J = \ell-1$ and (ii) series B modes with total angular momentum $J = \ell+1$. 
Each A and B mode is a superposition of $(\ell,s)$ and $(-\ell,-s)$ circularly-polarized modes, where $\ell>0$ and $s=-1$ for A modes and $s=+1$ for B modes. 
To distinguish between the + and - superposition within each set, we add a second component to the polarization label which we denote by $+$ and $-$, depending on the symmetry of the state under mirror action in the $xz$ plane.  
With these polarization and symmetry aspects in mind, we thus specify the vector quantum number as $v=\{y,x\}=\{+,-\}$ for the two $p,\ell=0$ modes and as $v=\{A+,A-,B+,B-\}$ for the four $p, \ell \geq 1$ modes.

\begin{figure}[tbh]
\centering
\includegraphics[width=\linewidth]{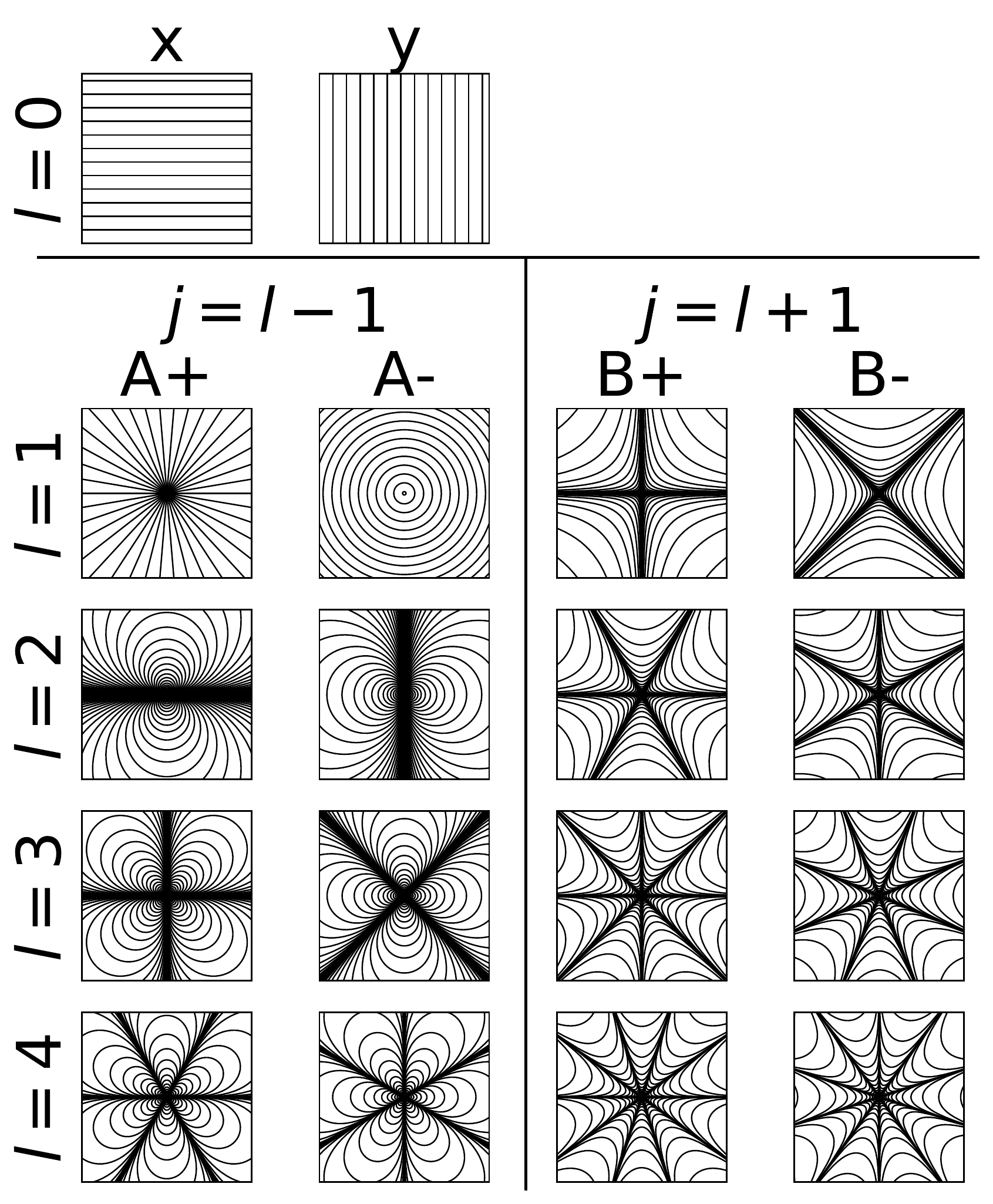}
\caption{Sketches of the polarization profiles, indicated as lines, of the vector LG-modes for $\ell = 0-3$. The two $\ell = 0$ modes are $x$-polarized and $y$-polarized. 
Each of the four $\ell \geq 1$ modes split in two A and two B modes; the 1A+ mode is radially polarized; the 1A- mode is azimuthally polarized; the other modes have a mixed radial/azimuthal polarization. 
Most modes occur in two versions, with field patterns that are rotated over half a lobe/finger with respect to each other. 
The +/- labels indicate whether the vector field is symmetric (+) or anti-symmetric (-) upon reflection in the $x$ axis. 
}
\label{fig:transverse-modes}
\end{figure}

As a vector generalization of the scalar paraxial modes of Eqs.~(\ref{eq:psi-pl})-(\ref{Eq:amplitude_functions}), we write the transverse field of the vector-LG modes as
\begin{equation}
\label{eq:psi-vector}
    \vec{\psi}_{p,\ell,v}^+(r,\theta,z) = \frac{1}{\gamma_z} \, \vec{e}_{\ell,v}(\theta)  f_{p,\ell}(\rho) \exp\left[ ik\frac{r^2}{2R_z} - i (N+1)\chi \right] \,. 
\end{equation}
For $\ell =0$ modes: $\vec{e}_{0,+}(\theta)=\vec{e}_x$ and $\vec{e}_{0,-}(\theta)=\vec{e}_y$.
For $\ell \geq 1$ modes, the vector fields are also linearly polarized, but the orientation of this linear polarization depends on $\theta$ as~\cite{Yu1983}
\begin{widetext}
\begin{equation}
\begin{array}{cccccc}
    \vec{e}_{\ell,v}(\theta) & = & \cos{(\ell \theta)} \vec{e}_x + \sin{(\ell \theta)} \vec{e}_y  & = & \cos{[(\ell-1) \theta]} \vec{e}_r + \sin{[(\ell-1) \theta]} \vec{e}_\theta \quad & (\rm{for} \,\, v=A+) \,, \\
    \vec{e}_{\ell,v}(\theta) & = & - \sin{(\ell \theta)} \vec{e}_x + \cos{(\ell \theta)} \vec{e}_y & = & -\sin{[(\ell-1) \theta]} \vec{e}_r + \cos{[(\ell-1) \theta]} \vec{e}_\theta \quad & (\rm{for} \,\, v=A-)  \,, \nonumber \\
    \vec{e}_{\ell,v}(\theta) & = & \cos{(\ell \theta)} \vec{e}_x - \sin{(\ell \theta)} \vec{e}_y  & = & \cos{[(\ell+1) \theta]} \vec{e}_r - \sin{[(\ell+1) \theta]} \vec{e}_\theta \quad & (\rm{for} \,\, v=B+)  \,, \nonumber \\
    \vec{e}_{\ell,v}(\theta) & = & \sin{(\ell \theta)} \vec{e}_x + \cos{(\ell \theta)} \vec{e}_y & = & \sin{[(\ell+1) \theta]} \vec{e}_r + \cos{[(\ell+1) \theta]} \vec{e}_\theta \quad & (\rm{for} \,\, v=B-)  \,. 
\end{array}
\end{equation}
\end{widetext}
Figure~\ref{fig:transverse-modes} shows the polarization profiles of the A and B modes for $\ell=0-3$. 
Note that all + modes have $\vec{e}(\theta=0) = \vec{e}_x$ and all - modes have $\vec{e}(\theta=0) = \vec{e}_y$.
Our short-hand notation for these modes is $|0X+\rangle$ and $|0Y-\rangle$ for the $\ell=0$ modes and $|\ell A \pm \rangle$ and $|\ell B \pm \rangle$ for the $\ell \geq 1$ $A\pm$ and $B\pm$ modes.
The radial dependence $f_{p\ell}(\rho)$ is not included in this labeling, but can be easily added with an extra quantum number $p$ or $N=2p+\ell$.

The twofold frequency degeneracy expected in a cavity with rotation and mirror symmetry has two consequences. 
First, it makes it difficult in experiments to find the true eigenmodes of the cavity, as the output field $| \vec{\psi}_{\rm out} \rangle$ will be a superposition of the two degenerate modes with relative amplitudes that are determined by the input field $| \vec{\psi}_{\rm in} \rangle$. 
Second, in the theory it leaves room for an alternative labeling of the vector modes. 
Zeppenfeld and Pinkse~\cite{Zeppenfeld2010} chose the total angular moment $J=\ell+s$ and the circular polarization $\sigma_\pm$ as labels, instead of our $\ell,v$ labels.
The vector profiles of their $(J,\sigma_\pm)$ modes, which are of the form $\vec{e}_{J,\sigma_\pm}(\theta) = \vec{e}_\pm \exp{\pm i \ell \theta}$ with $\vec{e}_\pm = (\vec{e}_x \pm i \vec{e}_y)$ as circular polarizations, are linear superpositions of our A and B modes (see Appendix~\ref{sec:Appendix-Zeppenfeld} and \ref{sec:Appendix-operators}).

\subsection{Non-paraxial vector correction ${\cal H}_{\rm vec}$}
\label{sec:vector-sub}

The non-paraxial vector correction ${\cal H}_{\rm vec}$ originates from the vector character of the optical field, and in particular from the small axial component of the optical field.
At the curved mirror, part of this axial field transforms into a reflected transverse field because the boundary condition is not decribed by $\vec{E}_\perp=\vec{0}$ but by the requirement that ``$\vec{E}$ is directed along the surface normal". 
Cullen~\cite{Cullen,CullenYu} was one of the first to mention this vector correction. 
Davis~\cite{Davis1984} quantified it for a linearly-polarized fundamental Gaussian mode, using geometric arguments. 
The vector correction to the reflection produces an effective spin-orbit coupling that is similar to the one observed under strong focusing \cite{Bliokh2010,Bliokh2015}.

Yu and Luk~\cite{Yu1984} and Luk~\cite{Luk1986} generalized the analysis to any vector LG-modes. 
They derived the vector corrections of these modes with the so-called action theorem, which is based on thermodynamic arguments and relates the relative frequency shift to the relative change in stored energy, via $\Delta f/f = \Delta W/W$~\cite{Davis1984,Yu1984,Luk1986,Uphoff2015}. 
In this subsection, we will instead use the roundtrip-matrix formalism to generalize the calculation to the vector coupling between any pair of modes. 
The roundtrip formalism is more general because it yields a coupling matrix, whereas the action theorem only yields the on-diagonal elements of this matrix.

We start our analysis with the earlier statement that every non-uniform transverse field has an axial component $\psi_z \approx (i/k) \vec{\nabla}_\perp \cdot \vec{\psi}_\perp$, see Eq.~(\ref{eq:longitudinal}).
For the transverse vector field described by Eq.~(\ref{eq:psi-vector}), the operator $\vec{\nabla}_\perp$ yields three contributions: 
The derivative of the phase factor in Eq.~(\ref{eq:psi-vector}) yields the in-phase field $\psi_z = - (r/R) \psi_\perp$ needed to orient the vector field of the traveling wave along the curved wavefront; this field has no further consequences. 
By contrast, the derivatives of the two other factors in the right-hand side of Eq.~(\ref{eq:psi-vector}), which together form the normalized derivative $\vec{\nabla}_\perp \cdot \vec{\Psi}(\rho,\theta)$, yield a small out-of-phase longitudinal field that projects into an additional  radially-polarized transverse field that does modify the resonance. 
This projection includes a geometric factor $- \vec{r}/R_m$ and a factor 2 to account for the standing-wave character of the field, similar the factor 2 in the phase lag $2k\Delta z$ that described the mirror shape.
The resulting additional transverse field is
\begin{equation}
\label{eq:project}
    \vec{\Psi}_{\perp, {\rm project}} = - i \frac{2\vec{\rho}}{kR_m} 
    \vec{\nabla}_\perp \cdot \vec{\Psi}_\perp(\vec{r}) 
    \,,
\end{equation}
where $\vec{\rho} = \vec{r}/\gamma_z$ and $\vec{\nabla}_\perp = \vec{e}_x (\partial/\partial \rho_x) + \vec{e}_y (\partial/\partial \rho_y) = (\vec{e}_r/\rho)(\partial/\partial \rho)\rho + (\vec{e}_\theta/\rho)\partial/\partial \theta$ is the transverse derivative vector operator in normalized coordinates. 
By comparing Eq.~(\ref{eq:project}) with the generic equations in Sec. \ref{sec:Roundtrip}, we find the non-paraxial vector correction 
\begin{equation}
\label{eq:vector-matrix}
    \Delta \tilde{\nu}_{{\rm vec},j',j} = \frac{1}{\pi k R_m} \langle \vec{\Psi}_{j'} | \, \vec{\rho} \otimes \vec{\nabla}_\perp | \vec{\Psi}_j \rangle \,,
\end{equation}
where the tensor product symbol $\otimes$ indicates that $\vec{\nabla}_\perp$ operates on $|\vec{\Psi}_j \rangle$ and $\vec{\rho}$ operates on $\langle \vec{\Psi}_{j'} |$. 

The rotation symmetry of the $\vec{\rho} \otimes \vec{\nabla}_\perp$ operator imposes conservation of total angular momentum $J'=J$.
In Appendix~\ref{sec:Appendix-operators} we will show that vector coupling is even diagonal, both in the basis of the vector-LG modes and in the basis of the $(J,\sigma_\pm)$ modes of Zeppenfeld of order $N$.
The corresponding (normalized) frequency shifts are 
\begin{equation}
\label{eq:vec}
    \Delta \tilde{\nu}_{{\rm vec},j} = \frac{\pm \ell -1 }{2\pi kR_m} = \frac{-1-\ell \cdot s}{2\pi kR_m} \,,
\end{equation}
where the + sign applies to $A$-modes, with total angular momentum $J=\ell-1$, and the - sign applies to $B$-modes, with $J=\ell+1$. 

Equation~(\ref{eq:vec}) agrees with earlier results presented in Refs.~\cite{Yu1984,Luk1986,Zeppenfeld2010}.
For $\ell \geq 1$ modes, the vector correction acts as an effective $L-S$ coupling, denoted by $\ell.s$ in Eq.~(\ref{eq:vec}), which splits each set of four $\ell \geq 1$ modes into two pairs of frequency-degenerate vector modes.
For the two, $x$ and $y$-polarized, $\ell=0$ modes, Eq.~(\ref{eq:vec}) yields equal shifts.
These results can be derived by the operator algebra described in Appendix~\ref{sec:Appendix-operators} or by partial integration over vector fields of the form $\vec{\Psi}_j(\rho,\theta) = \vec{e}_j(\theta) f_j(\rho)$.
For the $x$-polarized $\ell = 0$ mode, $| \vec{\psi} \rangle = \vec{e}_x | \psi \rangle$, the result simplifies to $\langle \vec{\Psi} | \vec{r} \otimes \vec{\nabla}_\perp | \vec{\Psi} \rangle = \langle \Psi | x \partial_x | \Psi \rangle = -1/2$.

\subsection{Combined result for rotation-symmetric cavities}

Let's combine the effects discussed up to now, into a single equation for the fine structure of a plano-concave cavity with a simple spherical mirror.
By combining Eqs. (\ref{eq:scalar-total}) and (\ref{eq:vec}), we find that the non-paraxial contribution to the roundtrip phase lag is
\begin{equation}
\label{eq:final-summary}
    \varphi_{\rm non} = 2\pi \Delta \tilde{\nu} = \frac{1}{kR_m} [ \frac{1}{8}(N^2+2N-4) - \frac{3}{8} \ell^2 - \ell \cdot s ] \,, \\
\end{equation}
This result is identical to the results of Refs.~\cite{Luk1986} and~\cite{Zeppenfeld2010}. 
For a plano-concave cavity with a non-spherical rotation-symmetric mirror, the factor $3/8$ in the top equation is replaced by $[3/8+\tilde{p}L/8(R_m-L)]$, due to spherical aberration and the polynomial in $N$ is also slightly modified; see Eqs.~(\ref{eq:asphere}) and~(\ref{eq:Zeppenfeld}).

\section{Bragg (vector) correction ${\cal H}_{\rm Bragg}$}
\label{sec:Bragg}

If we replace the ideal mirror by a more realistic Bragg mirror, the optical field will penetrate in the mirror and the reflection amplitude will change from a steady $r=-1$ to $r = - \exp[i\varphi(\omega,k_\perp)]$ \cite{Babic1992,Babic1993,Koks2021}.
The reflection phase $\varphi(\omega,k_\perp) = 2 k L_\varphi(\omega,k_\perp)$ will now depend on the optical frequency $\omega$ and the angle of incidence, which we express via its transverse momentum $k_\perp$ to avoid confusion with the orientation angle $\theta$.
The reflection phase is typically different for TE ($s$-polarized) light than for TM ($p$-polarized) light, because TE-light typically reflects better and thus yields a wider spectral stopband. 
The resulting phase difference $\varphi_s - \varphi_p = C (k_\perp/k)^2$ can impose an additional fine structure on the cavity modes and push the intra-cavity field towards radially- and azimuthally-polarized eigenmodes.
Foster et al.~\cite{Foster2009} have described this Bragg effect and compared its strength relative to the non-paraxial vector correction. 
We will briefly add our thoughts to their treatment.

First of all, we note that the phase difference $\varphi_s - \varphi_p = C (k_\perp/k)^2$, effectively also contains a $(k_\perp/k)^4$-term because $C$ depends on $(k_\perp/k)^2$ as $C(\vec{k}_\perp) = C_0 + C_2 (k_\perp/k)^2$ and $C_0 = 0$ in the center of the stopband~\cite{Koks2021}. 
Keeping this in mind, we write the perturbation imposed by the Bragg effect in the two mirrors as
\begin{equation}
\label{eq: hamiltonian bragg effect}
    {\cal H}_{\rm Bragg} = \frac{2C(k_\perp)}{k^2} \, \left( \vec{k}_\perp \otimes \vec{k}_\perp - \frac{1}{2} k^2_\perp \right) \,, 
\end{equation}
where the final $k^2_\perp/2$ balances the effect to zero for unpolarized light.
Appendix \ref{sec:Appendix-Bragg} calculates $C_0$ and $C_2$ from the properties of the DBR.
The $(k_\perp/k)^4$-term was not mentioned by F\"{o}ster et al.~\cite{Foster2009}, but could be relevant in experiments. 

The rotation symmetry of the Bragg operator matches the rotation symmetry of the vector LG-modes.
As a result, many matrix elements $\langle \psi_{j'}| {\cal H}_{\rm Bragg} | \psi_{j} \rangle$ are zero. 
The only non-zero on-diagonal elements of the Bragg operator are $J'=J=0$.
The Bragg operator will thereby split each pair of $J=0$ modes in a radially-polarized 1A+ mode and an azimuthally-polarized 1A- mode with opposite frequency shifts equal to
\begin{equation}
\label{eq:Bragg}
    \Delta \tilde{\nu}_{\rm Bragg,1A\pm} \approx \pm \frac{C_0}{kz_0}(N+1) \,.   
\end{equation}
To arrive at this result, we only included the $C_0$ term of the Bragg effect and used the mean-square opening angle $\langle k_\perp^2/k^2 \rangle = (N+1)/kz_0$ (see Appendix \ref{sec:Appendix-Bragg}). 
The relative strength of the predicted Bragg splitting, compared to the common factor $1/(8\pi kR_m)$ for non-paraxial effects, is $Y \equiv (8\pi kR_m)\Delta \tilde{\nu}_{\rm Bragg,1A+} \approx 16 \pi C_0 \sqrt{R_m/L}$ for $L \ll R_m$ and $N=1$.
Bragg effects become more prominent for cavities with $L \ll R_m$ because they scale with the mean-square modal opening angle, which increases when $L$ decreases. 
The observation of a 1A+/1A- splitting is a hallmark for the Bragg effect.

Dufferwiel et al.~\cite{Dufferwiel} have observed a TE-TM splitting between the $1A+$ and $1A-$ mode for a cavity filled with an active semiconductor.
They attributed the observed effect to TE-TM splitting of the polariton eigenstates associated with two different branches in the semiconductor band structure. 
But a remnant of the Bragg effect might also have been present.

As an aside, we note that the $\ell=1$ spectra shown in Ref.~\cite{Dufferwiel} actually comprises three peaks. 
The frequency difference between the outer $1A+$ and $1A-$ is due to the mentioned TE-TM splitting. 
But the average frequency of these peaks doesn't coincide with the frequency of the inner peak, which must originate from the degenerate $1B+$ and $1B-$ modes. 
We think that this additional frequency difference is due to the spin-orbit coupling described in our Sec.~\ref{sec:vector-sub}.

The Bragg operator also has non-zero off-diagonal elements that couple $(p-1,\ell=J+1,A)$ modes with $(p,\ell=J-1,B)$ modes, both of order $N=2p+J-1$ (see Appendices~\ref{sec:Appendix-operators} and \ref{sec:Appendix-hyperfine}).  
The effect of these off-diagonal elements on the fine structure is limited in rotational-symmetric cavities, because the on-diagonal elements typically differ a lot and dominate.
We thus expect hardly any Bragg-related hyperfine splitting for $J \neq 0$ modes in these cavities.
From a physics perspective, the $J \neq 0$ vector modes are (to first order) insensitive to the Bragg effect because they contain equal amounts of radial and azimuthal polarization.
But even $J \neq 0$-type vector modes can exhibit some Bragg-induced hyperfine splitting when the cavity is sufficiently astigmatic to mix LG-modes, c.q. modify the eigenmodes, and thereby make them sensitive to off-diagonal matrix elements and the Bragg effect (see Appendix \ref{sec:Appendix-hyperfine}).

\section{Mirror-astigmatic corrections}
\label{sec:mirror}
\subsection{Astigmatic correction ${\cal H}_{\rm astigm}$}
\label{sec:astigm}

In the previous sections we analyzed the resonances in a plano-concave cavity with a rotationally symmetric and almost spherical mirror.
In this section we will analyze mirror deformations that lack rotation symmetry, which are known as astigmatic deformations.
Many concave mirrors are not rotationally symmetric but have slightly different curvatures in two orthogonal directions, which we will call $x$ and $y$. 
We describe the (paraboloidal component of the) astigmatic mirror shape as
\begin{equation}
    z_{\rm mirror}(x,y) = \frac{x^2}{2R_x} + \frac{y^2}{2R_y} \approx \frac{x^2+y^2}{2\overline{R}} + \eta_{\rm astigm} \frac{x^2-y^2}{2\overline{R}}  \,,
\end{equation} 
where $\overline{R}=(R_x+R_y)/2=R_m$.
The parameter $\eta_{\rm astigm}=(R_y-R_x)/(2\overline{R}) \ll 1$ quantifies the strength of the astigmatism. 

Astigmatism breaks the rotation symmetry and prefers Hermite-Gaussian modes over Laguerre-Gaussian modes. 
A combined treatment of astigmatism and rotationally-symmetric perturbation thus requires a matrix description that includes all relevant modes.
As astigmatism also has mirror symmetry in the (just-defined) $x$ axis, it only couples modes with the same +/- vector character.
Each $N$ group is thus expected to split in two subgroups, the $N+$ group and $N-$ group.

If astigmatism would be the only effect, then we would simply use the HG-modes as a basis instead of the LG-modes.
Based on the factorization of the HG profiles, we would then conclude that (i) the HG-modes are the eigenmodes of the astigmatic cavity and (ii) the modes in each $N$-group split into (pairs of) vector HG-modes with an equidistant spacing $\propto \eta_{\rm astigm}$. 
But this scalar analysis does not take spin-orbit coupling into account. 
To include this effect, we will instead analyze astigmatism in the basis of the vector LG-modes.

We start with the $N=1$ group.
This group contains four modes and splits into two pairs: the $(A+,B+)$ set and the $(A-,B-)$ set.
The astigmatic coupling matrix for each of these sets is (see Appendix~\ref{sec:Appendix-operators}) 
\begin{eqnarray}
    \Delta \tilde{\nu}_{{\rm astigm}, (N=1)} = \begin{pmatrix} 0 & \tilde{X} \\ \tilde{X} & 0 \end{pmatrix} \,, \\
\label{eq:tilde-X}
        \tilde{X} \equiv \frac{2}{\lambda} \langle 1A+ | \Delta z(x,y) | 1B+ \rangle =  \eta_{\rm astigm} \frac{\tan{\chi_0}}{2\pi} \,.
\end{eqnarray}
This result was checked via integration, using the relation $\langle \psi | r^2 | \psi \rangle = (w_1^2/2)(N+1)$ with $w_1^2 = (\lambda R/\pi) \tan{\chi_0}$.  
The eigenvalues of this astigmatic matrix are $\lambda_\pm = \pm \tilde{X}$.
The associated eigenmodes, $(A+) \pm (B+)$, have a $\cos{\theta}\vec{e}_x$ and $-\sin{\theta}\vec{e}_y$ angular dependence. 
They are the mirror-symmetric vector $HG_{10}$ and $HG_{01}$ modes that one expects in the absence of spin-orbit coupling. 

Next, we add the non-paraxial corrections as on-diagonal elements $\Delta \tilde{\nu}_j = [g(p,\ell) - 4 \ell \cdot s] / (8 \pi k R_m)$ with $g(0,1)=-2$.
We also introduce the relative strength of the astigmatism as 
$X = (8\pi k R_m) \tilde{X} = 2 \eta_{\rm astigm} k^2 w_1^2$ to remove a common factor.
The final result in the $(1B+,1A+)$ basis 
\begin{equation}
\label{eq:N1}
    (8\pi k R_m)\, \Delta \tilde{\nu}_{(N=1)} = \begin{pmatrix} -6 & X \\ X & 2 \end{pmatrix} 
\end{equation}
combines astigmatism with non-paraxial corrections. 
The eigenvalues of this complete matrix are $\lambda_{\pm} = -2 \pm \sqrt{16+X^2}$.
The new eigenmodes are of the form $[\cos{\beta}\psi_A+\sin{\beta}\psi_B]$ and $[-\sin{\beta}\psi_A+\cos{\beta}\psi_B]$, with the mode-mixing angle $\beta = \arctan{(X/4)}$. 
This shows that astigmatism increases the splitting between the non-paraxial modes by a factor $\sqrt{1+(X/4)^2}$, while gradually changing the $A+$ and $B+$ eigenmodes that are visible at $X=0$ into the $x$-polarized $HG_{10}$ and $y$-polarized $HG_{01}$ eigenmodes for $X \gg 1$.

To calculate the astigmatic matrix for any $N \geq 2$ group, we need to find the associated matrix elements. The $x^2-y^2 = r^2 \cos{2\theta}$ angular dependence of the perturbation, shows that the astigmatic coupling obeys the selection rule $\Delta \ell = \pm 2$.
Furthermore, astigmatism only couples $\ell \leftrightarrow (\ell+2)$ modes with the same vector label $v = \{A+,A-,B+,B-\}$. 
But astigmatism can also couple $1A$ and $1B$ modes, because the $\ell=1$ modes implicitly also contain $-1$ modes.
Hence, the only non-zero matrix elements of the astigmatic contribution are
\begin{eqnarray}
\label{eq:tilde-X1}
        \frac{2}{\lambda} \langle \psi_{p,\ell=1,A+} | \Delta z(x,y) | \psi_{p,\ell=1,B+} \rangle & = &  \tilde{X} (p+1) \,, \\
        \frac{2}{\lambda} \langle \psi_{p+1,\ell-2,v} | \Delta z(x,y) | \psi_{p,\ell,v} \rangle & = & \tilde{X} h(N,\frac{N-\ell}{2}) \,, \\
        \frac{2}{\lambda} \langle \psi_{p-1,\ell+2,v} | \Delta z(x,y) | \psi_{p,\ell,v} \rangle & = & \tilde{X} h(N,\frac{N+\ell}{2}) \,, 
\end{eqnarray}
were $v$ can be any vector label. 
The first equation has an identical counterpart for the $-$ modes.
The second equation assumes $\ell \geq 2$ and introduces the function $h(N,n_s) \equiv \sqrt{(n_s+1)(N-n_s)}$, which obeys the symmetry $h(N,N-n_s)=h(N,n_s-1)$.
These results were again obtained with the operator algebra described in Appendix \ref{sec:Appendix-operators}. 


\begin{figure}
    \centering
    \includegraphics[width=\linewidth]{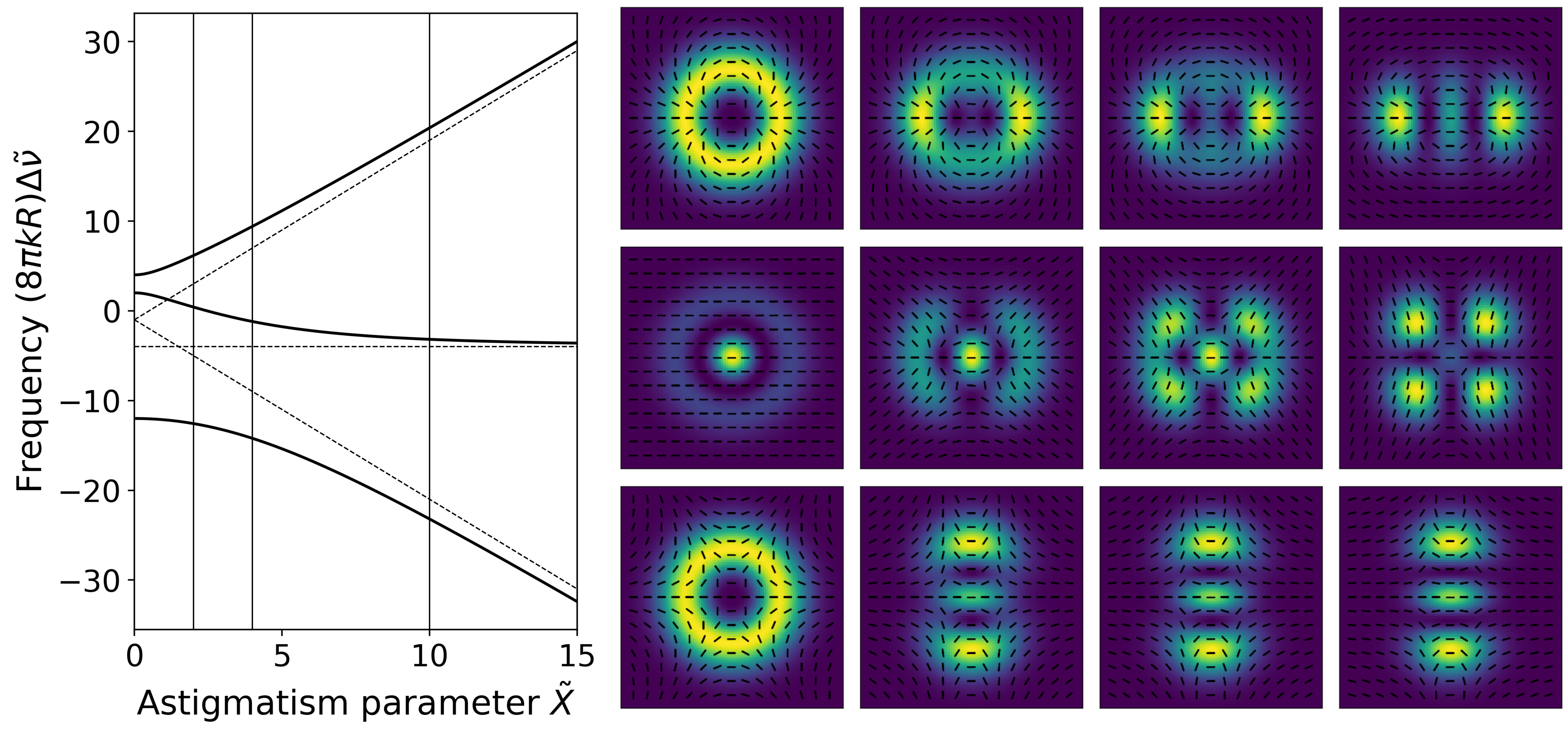}
    \caption{Sketches of the polarisation profile for the "+" modes of the $N=2$ group, as a function of the normalized astigmatism $X$. The eigenmodes on the right correspond to $X=0, 2, 4, 10$.}
    \label{fig:astigmatism}
\end{figure}

Using the results presented above, we can now calculate the coupling matrix for any non-paraxial astigmatic cavity. 
We will show the result only for the $N=2$ and $N=3$ groups and leave the general analysis to the reader. 
The $N=2$ group contains $2(N+1)=6$ members, 3 with a $+$ mirror symmetry and 3 with a $-$ mirror symmetry. 
The three LG-vector modes with a $+$ character are the $(0+,2A+,2B+)$. 
The on-diagonal elements of the spectral matrix are determined by the non-paraxial correction $\left[ g(p,\ell) \pm 4\ell \right] = 2$ for the $0$-mode, $4$ for the $2A$-mode and $-12$ for the $2B$-mode.
The off-diagonal elements are determined by the astigmatism, which couples the $\ell =0$ mode with the two $\ell =2$ modes with an equal normalized coupling $\sqrt{2}X$. 
This makes the combined spectral matrix in the $(2B+,0+,2A+)$ basis equal to
\begin{equation}
\label{eq:N2}
    (8\pi k R_m)\, \Delta \tilde{\nu}_{(N=2)} = \begin{pmatrix} -12 & \sqrt{2}X & 0 \\ 
    \sqrt{2}X & 2 & \sqrt{2}X \\ 0 & \sqrt{2}X & 4  \end{pmatrix} \,.
\end{equation}
The three $-$ modes are coupled by an identical matrix that now operates in the $(2B-,0-,2A-)$ basis.
A similar exercise for the $N=3$ group, where the set of $+$ modes are $(3B+,1B+,1A+,3A+)$, yields 
\begin{equation}
    (8\pi k R_m)\, \Delta \tilde{\nu}_{(N=3)} = \begin{pmatrix} -20 & \sqrt{3}X & 0 & 0 \\  \sqrt{3}X & 0 & 2X & 0 \\ 0 & 2X & 8 & \sqrt{3}X \\ 0 & 0 & \sqrt{3}X & 4  \end{pmatrix} \,.
\end{equation}

Figure~\ref{fig:astigmatism} shows the fine structure in the $N=2$ group as a function of the normalized astigmatism $X$ and the associated eigenmodes. 
In the absence of astigmatism, at $X=0$, the vector-LG $2A+$ and $2B+$ modes have eigenfrequencies $(8\pi k R_m) \tilde{\nu} = 4$ and -12, respectively, while the $0A+$ mode has eigenfrequency 2.
At non-zero astigmatism, these three modes mix and gradually transform from LG- to HG-modes, while their eigenfrequencies also change.
At $X = 10$, where astigmatism dominates over non-paraxial effects, the eigenmodes strongly resemble the HG-modes and the distance between the eigenvalues becomes approximately equal. 
The three asymptotes show the eigenfrequencies -4 and $-1 \pm 2X$ that are reached at $X \gg 1$. 
The depicted transition from dominant non-paraxial effects to dominant astigmatic effects in an optical cavity resembles the transition from dominant spin-orbit coupling to a dominant Zeeman effect in atomic physics. 

\subsection{Shape birefringence}
\label{sec:shape}

In the previous section, we stated that the coupling matrices of the $+$ and $-$ modes are identical. 
We will now show that these matrices can be slightly different on account of a second-order effect that combines astigmatism with the non-paraxial vector correction. 
The associated anisotropic spin-orbit coupling results in shape birefringence, i.e. it induces a frequency difference between $x$- and $y$-polarized light, just as birefringence would, but only because of the $x/y$ difference in the shape of a mirror. 
The story is as follows:

The vector correction in Eq.~(\ref{eq:project}) uses the transverse derivative $\vec{\nabla}_\perp . \vec{\Psi}_\perp(\vec{r})$ to calculate a small additional  field, which is then projected onto a radial transverse field by multiplying it with $\vec{\rho}/R_m$. 
But in a cavity with an astigmatic mirror, the projection should instead be on 
\begin{equation}
    \frac{\vec{\rho}_x}{R_x} + \frac{\vec{\rho}_y}{R_y} = \frac{\rho}{\overline{R}} \vec{e}_r + \eta_{\rm astigm} \frac{\rho}{\overline{R}} ( \cos{\theta} \vec{e}_x - \sin{\theta} \vec{e}_y ) \,.
\end{equation}
The first term in this equation describes the vector correction discussed in Sec. \ref{sec:vector}.
The second term described the astigmatic component of this vector correction.

The anisotropic vector correction can again be calculated via integration or operator algebra. 
For $\ell=0$ modes, the resulting on-diagonal matrix elements yield
\begin{equation}
\label{eq:HG00-splitting}
    \Delta \tilde{\nu}_{v+a} = \mp \frac{\eta_{\rm astigm} }{2 \pi k \overline{R}} \,, 
\end{equation}
where the $-$ sign applies to the $+$ mode and vice versa and where Eq.~(\ref{eq:HG00-splitting}) applies to all $\ell =0$ modes, irrespective of $p$.
The same result can be obtained by interpreting the vector correction of Eq.~(\ref{eq:vec}) as $\Delta \tilde{\nu} = - 1/(2\pi k R_x)$ for the $x$-polarized mode with $R_x = \overline{R}(1+\eta_{\rm astigm})$.
For $\ell \geq 1$ modes, the anisotropic spin-orbit coupling only has non-zero off-diagonal elements (see Appendices \ref{sec:Appendix-operators} and \ref{sec:Appendix-hyperfine}).
As a result, these modes experience typically hardly any shape birefringence and retain their $+/-$ degeneracy in cavities with small astigmatism. 
Some hyperfine splitting might, however, still be present in cavities where the astigmatism is strong enough to mix the vector LG-modes.

\section{Discussion \& residual ${\cal H}_{\rm rest}$}
\label{sec:Discussion}

The analysis presented above describes the most common perturbations in optical cavities, but is unavoidably incomplete.
We included the quartic scalar corrections $k_\perp^4$ and $r^4$, the spin-orbit and Bragg vector correction, and astigmatic deformations of the form $(x^2-y^2)$, but the concave mirror might also be deformed in different ways. 
We will combine all residual mirror deformations in the scalar operator ${\cal H}_{\rm rest} = 2 k \Delta z_{\rm rest} \ll 1$.

The residual operator ${\cal H}_{\rm rest}$ will scatter light and couple modes, just like the other operators do. 
And this coupling will again mainly be effective between modes with the same transverse order $N$, if the cavity is operated far from frequency degenerate points. 
The effect of ${\cal H}_{\rm rest}$ on the spectral fine structure and the associated eigenmodes is then completely described by the residual spectral matrix $\Delta \tilde{\nu}_{j',j} = \langle \psi_{j'} | {\cal H}_{\rm rest} | \psi_j \rangle$. 
In principle, knowledge of the spectrum and eigenmodes of a specific order $N$ allows one to reconstruct the full coupling matrix ${\cal H}_{\rm fine}$ of that order and disentangle its contributions. 
The accuracy of such an analysis is only limited by the cavity finesse, which makes it an extremely sensitive probe of the actual mirror shape.  
Benedikter et al. \cite{Benedikter2019} have previously used the resonances around frequency degenerate points as a similar sensitive probe for the topography of their planar mirror. 

The eigenmodes of the spectral matrix will provide a better match with the deformed mirror than the original LG-modes, but the match is typically not perfect.
The resulting modal loss per roundtrip can be calculated from the next term in the Taylor expansion of  $\exp(-2ik\Delta z) \approx 1 - i 2k\Delta z - 2k^2\Delta z^2$. 
The calculated amplitude loss
\begin{equation}
\label{eq:loss}
    \gamma_{{\rm extra}} \approx 2k^2 \iint \mbox{d}x \mbox{d}y \, [\Delta z(x,y)]^2 \, |\psi(x,y)|^2  \,,
\end{equation}
with $\Delta z = z_{\rm mirror} - z_{\rm mode}$, yields the famous expression $(4\pi\sigma/\lambda)^2$ with $\sigma^2 = \langle \Delta z^2 \rangle$ for the intensity loss~\cite{Harvey2012}.
A finite mirror radius $r_{\rm max}$ will add clipping amplitude loss 
\begin{equation}
\label{eq:clipping}
    \gamma_{{\rm clipping}} \approx \int_{r_{\rm max}}^\infty \pi d r^2 \, |\psi(r)|^2  \,,
\end{equation}
where $|\psi(r)|^2$ is the rotational-averaged intensity profile. 

At first sight, it might be surprising that height variations introduce modal loss in a system that was assumed to be lossless.  
But any submatrix ${\cal H}$ of the system is bound to be non-Hermitian even when the full dynamic matrix is Hermitian.
The residual loss originates from coupling between modes with different order $N$, which will increase the mode size beyond reasonable bounds and thereby result in clipping loss.
Even when the coupling between individual modes is inefficient, the multitude of available coupling channels creates the residual loss described by Eq.~(\ref{eq:loss}).

\section{Summary \& outlook}
\label{sec:Summary}

This paper calculates the resonance frequencies and eigenmodes of a planar-concave cavity beyond the common paraxial limit and beyond the spherical mirror shape. 
It does so by describing the roundtrip dynamics of the intra-cavity field in a 
3-dimensional Fabry-Perot (FP) cavity in a general operator formalism, which reduces to a modest-size matrix description after the application of perturbation theory. 
It then shows that the $2(N+1)$ modes with the same longitudinal $q$ and transverse order $N$ are not frequency degenerate, as predicted by a paraxial theory, but split.
The associated optical fine structure has many contributions, which are listed in Table~I and calculated in parts~II and~III of the paper. 

The paper presents a complete theoretical framework for the expected optical fine structure, by systematically analyzing all contributions that can realistically be expected.
For cavities with rotation and mirror symmetry we basically recover the results of Refs.~\cite{Luk1986} and \cite{Zeppenfeld2010}, albeit often in easier forms.
This analysis is presented in part II, Secs. \ref{sec:scalar} and \ref{sec:vector}.
In part III, we analyze the fine structure of more general cavities. 
This results in four additional contributions: (i) a Bragg correction, to quantify the role of the Bragg mirror, (ii) an astigmatic correction to quantify the effect or an astigmatic mirror, (iii) an isotropic spin-orbit coupling in astigmatic cavities, and (iv) a residual correction, which was only discussed in general terms.  
In experiments, the astigmatic correction is expected to be an important technical complication that can easily dominate the more fundamental non-paraxial corrections. 
The paper introduces a dimensionless parameter $X$ to compare these effects and predicts how the fundamental effects are more likely to dominate in short cavities with mirrors with small radii of curvature.
A second dimensionless parameter $Y$ compares the strength of the Bragg effect relative to the more fundamental non-paraxial corrections.

The optical fine structure in FP spectra resembles the fine structure in atomic spectra. 
The energy levels in atomic physics depend primarily on the principal quantum number $n$, but exhibit a fine structure that is an order $\alpha^2$ smaller, where $\alpha \approx 1/137$ is the fine structure constant \cite{Griffiths}. 
In comparison, the optical resonances in FP spectra are primarily determined by the two principal quantum numbers $(q,N)$.
The relative strength of the optical fine structure $\Delta \tilde{\nu}$ is of the order $1/(8\pi k R_m) \propto \lambda/R_m$, where $R_m$ is the radius of curvature of the mirror. 

The results presented in Table~I describe the non-paraxial effects in a plano-concave cavity with mirror spacing $L$ and mirror radius $R_m$.
This result can easily be generalized to an arbitrary cavity, with two mirrors with radii $R_1$ and $R_2$, using the following procedure: (i) use paraxial optics to find the waist and the distances $L_1$ and $L_2$ from the mirrors to this waist, (ii) use Table~I to determine the various contributions to $\varphi_{\rm non,1}$ for the optical path from the waist to Mirror~1 and back, (iii) repeat this step for Mirror~2 to find $\varphi_{\rm non,2}$, and (iv) add the two results to find the non-paraxial roundtrip phase lag $\varphi_{\rm non} = \varphi_{\rm non,1} + \varphi_{\rm non,2}$.
For a symmetric bi-concave cavity of length $2L$, one thus finds that (i) $\varphi_{\rm non}$ is twice as large as in a plano-concave cavity of length $L$, (ii) the relative frequency shift $\Delta \tilde{\nu}$ is also twice as large, but (iii) the absolute frequency shifts $\Delta \nu$ are equal, as expected. 

The analysis presented in this paper neglects losses. 
This assumption is valid for cavities with large high-reflective mirrors, such that the cavity resonances are clearly resolved in the optical spectrum. 
The influence of loss on the spectral resonances is probably limited to detunings $\Delta \tilde{\nu} < 1/F$ and therefore small for finesses $F \gg 10 R/\lambda$, where the factor 10 was added to compensate for a factor $\ll 10$ in Eq. (\ref{eq:final-summary}). 
A measurement of the modal finesses, and the associated clipping losses at finite-size mirrors, can however be useful to further characterize the individual modes.

The analysis predicts that most modes appear in frequency-degenerate pairs, with polarization patterns of the form $A\pm$ or $B\pm$ for the $\ell \geq 1$ modes and $x$ and $y$ polarization for the $\ell = 0$ modes. 
It also predicts that this pairwise degeneracy will be slightly broken for some pairs by effects that one might thus call hyperfine splitting.
This paper quantifies two effects and shows that: (i) the $\ell=1,A$ pairs split in modes with radial and azimuthal polarization in cavities with Bragg mirrors and (ii) the $\ell = 0$ mode pairs exhibit a small second-order splitting in astigmatic cavities.
It also argued how the degeneracy of other mode pairs could be slightly broken in strongly-astigmatic cavities, due to admixture of HG-character in the vector LG-modes that are preferred by spin-orbit coupling. 
The pairwise degeneracy can probably also be broken when the mirror symmetry is broken, for instance when one mirror has a twist~\cite{Blows,Kasutich}, or a higher-order astigmatism of the form $x'^4-y'^4$ with an $x'y'$ orientation different from the $xy$-orientation of the prime astigmatism.
And mirror symmetry is obviously broken in cavities with chiral stuctures, like the ones recently reported in Ref. \cite{Gautier2021} 

As a further outlook, we note that the optical fine structure contains information on the mirror shape down to sub-nm precision. 
It can thus in principle be used to inspect these shapes, without the need to dismount the mirrors and inspect them by AFM or optical interference. 

In future work, the analysis could be extended by including the other, $C_2$ or $k_\perp^4$, Bragg effect.
This effect was neglected in most of the analysis, but the example presented in Appendix \ref{sec:Appendix-Bragg} shows that this simplification is not always correct.

The analysis could also be extended by including the coupling between modes of different $N$-groups. 
The latter coupling is typically small but will become important at so-called frequency-degenerate cavity lengths, where the Gouy phase $\chi_0$ is a rational fraction of $\pi$ and modes with different $(q,N)$ numbers become frequency degenerate.
The resulting modified eigenmodes can potentially lead to a reduction in mode area and an increase the light-matter interaction~\cite{Koks2022a}.
The analysis could also be extended to optical cavities with different geometries, beyond the two-mirror plano-concave type.
The theoretical framework developed in this paper is general enough to also analyze these related geometries in a perturbative way. 

Finally, it might be interesting to compare the presented analysis with the geometric approach to cavity aberration presented in a recent publication of Jaffe et al. \cite{Jaffe2021}.
Or to compare the presented analysis of mode formation in open optical cavities with the analysis of mode formation in rotational-symmetric graded-index optical waveguides/fibers presented in ref. \cite{Petrov2013}.
We leave these topics as challenges to the reader. 

\begin{acknowledgments}
The authors would like to thank Sean van der Meer and Martin Bijl for experiments that stimulated our ideas on the optical fine structure.
They also acknowledge Gerard Nienhuis for his suggestion to investigate the operator formalism \cite{Nienhuis2004,Nienhuis} and Pepijn Pinkse for sharing his knowledge on prior experiments on the cavity fine structure \cite{Zeppenfeld-CLEO}. 
MW acknowledges support by the Danish National Research Foundation through NanoPhoton - Center for Nanophotonics, grant number DNRF147 and Center for Nanostructured Graphene, grant number DNRF103, and by the 
Independent Research Fund Denmark - Natural Sciences (project no. 0135-00403B).
\end{acknowledgments}

\appendix
\section{Comparison with Zeppenfeld-Pinkse}
\label{sec:Appendix-Zeppenfeld}

This Appendix summarizes the key results of the article entitled 'Calculating the fine structure of a Fabry-Perot resonator using spheroidal wave functions' by Zeppenfeld and Pinkse~\cite{Zeppenfeld2010}, and compares them with the associated non-paraxial effects listed in Table~I.
We will transform their results to our notation for the specific case of a plano-concave cavity.

Zeppenfeld and Pinkse~\cite{Zeppenfeld2010} label their vector-LG modes with quantum numbers $(\nu,J,\sigma_\pm)$.
Zeppenfeld's quantum number $\nu$ is equal to our radial quantum number $p$.
Zeppenfeld's quantum number $J$ denotes the total angular momentum. 
As such, it combines our radial quantum number $\ell$ with a vector quantum number $s$ that is $\pm 1$ for circular-polarized $\sigma_\pm$ light.
Their $v=\{J,\sigma_+\}$ mode is a superposition of our $B_+$ and $B_-$ mode with $\ell=J-1$.
Their $v=\{J,\sigma_-\}$ mode is a superposition of our $A_+$ and $A_-$ mode with $\ell=J+1$.


Equation~(40) of Ref.~\cite{Zeppenfeld2010} states that the roundtrip phase of the $(\nu,J,\sigma^+)$ mode is 
\begin{widetext}
\begin{equation}
\label{eq:Zeppenfeld}
    \varphi_{\nu,J,\sigma^+} = 2kL - 2 (2\nu+J) \arctan{\xi_+} - \frac{2}{kR_m} \nu (\nu+J) + \frac{\xi_+}{kz_0} (\frac{1}{4} - \frac{L}{4R_m}\tilde{c}_4) [6\nu(\nu+J)+J(J+1)] \,, \\
\end{equation}
\end{widetext}
To arrive at Eq.~(\ref{eq:Zeppenfeld}), we combined Zeppenfeld's expansion parameter $1/\overline{c}=2/(kd)$, with $d=2z_0$, with equations that are specific for plano-concave cavities, like $\xi_+=2L/d$, $\xi/[\overline{c}(1+\xi^2)] =1/(kR_m)$, $f_{4+}=(2R_m^2L\tilde{c}_4-\frac{1}{4})\xi$, and Rayleigh range $z_0=\sqrt{L(R_m-L)}$.
The parameter $\tilde{c}_4$ describes the deviation from a paraboloidal mirror shape and is related to the parameter $\tilde{p}$ in the main text via $\tilde{c}_4 = 1 -\tilde{p}$.

The first term on the right-hand side of Eq.~(\ref{eq:Zeppenfeld}) is the plane-wave roundtrip phase. 
The second term is the phase lag predicted by paraxial theory.
The third and fourth terms in Eqs.~(\ref{eq:Zeppenfeld}) describe the spectral fine structure in Zeppenfeld's notation. 
We transform this to our notation by writing $J = \ell + s$ for $\ell >0$, with $J=\ell+1$ for $\sigma_+$ polarization, $2\nu+J = N+1$ with transverse order $N=2p+\ell$, and by introducing the fundamental Gouy phase $\chi_0 = \arctan{\xi} = \arctan {\sqrt{L/(R_m-L)}}$. 
With this rewrite, we arrive at the equivalent equation in our notation
\begin{eqnarray}
\label{eq:Zeppenfeld2}
    \varphi_{p,\ell,\sigma^\pm} & = & 2kL - 2 (N+1) \chi_0 - \varphi_{\rm non}\,,\quad\mbox{with} \nonumber \\
    \varphi_{\rm non} & = & \frac{1}{kR_m} [ \frac{1}{8}(N^2+2N-4) - \frac{3}{8} \ell^2 - \ell \cdot s ] \,.
\end{eqnarray}
This final result includes a rewrite of Eq.~(41) of Ref.~\cite{Zeppenfeld2010} and is therefore valid for modes with both types of polarization. 
It only applies to a plano-concave cavity with spherical mirrors ($\tilde{c}_4=1$ in Zeppenfeld's notation), but can be extended to aspherical mirrors by replacing the factor $3/8$, in front of the $\ell^2$-term, by $3/8+\tilde{p}L/[8(R_m-L)]$ and by slightly modifying the function of $N$.
Equation~(\ref{eq:Zeppenfeld2}) is identical to Eq.~(\ref{eq:final-summary}) in the main text, and to the results of Luk~\cite{Luk1986}. 

\section{Bragg correction in detail}
\label{sec:Appendix-Bragg}

This Appendix analyzes the polarization dependence of the reflection phase of a DBR and the resulting Bragg correction ${\cal H}_{\rm Bragg}$. 
The reflection phase of a DBR 
\begin{equation}
    \varphi_{s,p}(\omega,\phi) = 2 k L_\varphi(\omega,\phi) = [\omega - \omega_c(\phi)]\tau_{s,p}(\phi) 
\end{equation}
depends on the detuning $[\omega - \omega_c(\phi)]$ between the optical frequency and the center of the stopband and on the (polarization-dependent) phase penetration depth $L_{s,p}(\phi) = c\tau_{s,p}(\phi)/2$ in the DBR.
Both quantities depend on the angle of incidence $\phi \approx k_\perp/k \ll 1$ as~\cite{Babic1992,Babic1993}
\begin{eqnarray}
\label{eq:C2}
    \omega_c(\phi) & \approx & \omega(0) (1+B\phi^2) \,\, ; \, B=\frac{1}{4}(\frac{1}{n_L^2} + \frac{1}{n_H^2}) \,, \\
\label{eq:C3}
    \tau_{s,p}(\phi) & \approx & \tau(0) (1 \pm A\phi^2) \,\, ; \, A=\frac{1}{2}(1 + \frac{1}{n_Hn_L}) \,,
\end{eqnarray}
where the + sign in Eq. (\ref{eq:C3}) applies to $p$-polarized light, with its reduced Fresnel reflection and reduced stopband, and the - sign applies to $s$-polarized light.
Equation (\ref{eq:C2}) is the generic result of a Taylor expansion of Snell's law.
Equation (\ref{eq:C3}) is valid only for H-DBRs, i.e. DBRs that start with the high-index $n_H > n_L$ layer on the air side $(n_{\rm in}=1)$.  
By combining these expansions with the H-DBR result $\omega_c(0)\tau(0) = \pi/(n_H-n_L)$ \cite{Babic1992,Babic1993,Koks2021}, we find
\begin{equation}
\label{eq:p-s}
    \varphi_p(\phi) - \varphi_s(\phi) = \frac{2\pi}{n_H - n_L}\, A \phi^2 \, \left[ \frac{(\omega-\omega_c(0))}{\omega_c(0)} - B \phi^2 \right] \,.
\end{equation}

This polarization-dependent reflection at $k_\perp \neq 0$ creates the Bragg correction introduced in Eq.~(\ref{eq: hamiltonian bragg effect}) in the main text, which reads
\begin{equation}
\label{eq:H-Bragg}
    {\cal H}_{\rm Bragg} = \frac{2C(k_\perp)}{k^2} \, \left( \vec{k}_\perp \otimes \vec{k}_\perp - \frac{1}{2} k^2_\perp \right) \,, 
\end{equation}
with $C(\vec{k}_\perp) = C_0 + C_2 (k_\perp/k)^2$ and
\begin{equation}
    C_0 = \frac{2\pi A}{n_H - n_L} \frac{\omega-\omega_c(0)}{\omega_c(0)} \,\, ; \,\,
    C_2 = \frac{- 2\pi A B}{n_H - n_L} \,.
\end{equation}
The factor 2 in Eq. (\ref{eq:H-Bragg}), which indicates that there are two reflections, is only approximately 2, because the angle-dependent reflection from the curved mirror and flat mirror are only similar in the short-cavity limit $L \ll R_m$.
The ``quadratic Bragg effect", quantified by $C_0$, depends critically on the frequency detuning.
The ``quartic Bragg effect", quantified by $C_2$, does not and could thus even become dominant around the center of the stopband.
The ``quadratic Bragg effect" has on-diagonal elements only for the $j=j'=1A\pm$ modes, where
\begin{equation}
    \langle \psi_j | {\cal H}_{\rm Bragg} | \psi_j \rangle = \pm \frac{C_0}{kz_0} (N+1) \,,
\end{equation}
and where we used $\langle \phi^2 \rangle = \langle k_\perp^2/k^2 \rangle = (N+1)/kz_0$.

For a typical DBR coating of SiO$_2$  ($n_L \approx 1.46$) and Ta$_2$O$_5$ ($n_H \approx 2.09$), and a typical relative detuning $[\omega-\omega_c(0)]/\omega_c(0)$ of 1\%, we find $C_0 \approx -0.13$, while $C_2 \approx 2.3$ at any detuning. 
To calculate the associated polarization shifts, these values should be multiplied by $\langle \phi^2 \rangle = \langle k_\perp^2/k^2 \rangle = (N+1)/kz_0$ and $\langle \phi^4 \rangle = f(p,\ell)/(kz_0)^2$. 
For a typical microcavity with $L=2~\mu$m, $R = 20~\mu$m and $\lambda = 0.63~\mu$m, the rms opening angle of the fundamental mode $\sqrt{\langle \phi^2 \rangle} \approx \sqrt{0.016} \approx 0.13$ rad. 
For the $1A+$ and $1A-$ modes, this results in frequency shifts $\Delta \tilde{\nu}$ of $\pm C_0 \langle \phi^2 \rangle /(2\pi) = \mp 6.6 \times 10^{-4}$ due to the 1\% detuning and $C_2 \langle \phi^4 \rangle /(2\pi) = \pm 5.6 \times 10^{-4}$ due to the quartic correction.
These numbers show that the Bragg correction is typically small at small detuning, where the quartic effect typically also plays a role.
But the Bragg correction should be observable, in particular at larger frequency detuning.

For L-DBRs, i.e. DBRs that start with an $n_L$ layer on the air side, two parameters are different \cite{Babic1992,Babic1993,Koks2021}.
The product $\omega_c(0)\tau(0) = n_L n_H \pi/(n_H-n_L)$ is larger than for H-DBRs, but the polarization factor in Eq. (\ref{eq:C3}) is now
\begin{equation}
    A_L = \frac{1}{2}(\frac{1}{n_H^2} + \frac{1}{n_L^2} + \frac{1}{n_Hn_L} -1) \,,
\end{equation}
and is typically smaller.
For L-DBRs we thus find the modified equations, 
\begin{equation}
    C_0 = \frac{2\pi \tilde{A}}{n_H - n_L} \frac{\omega-\omega_c(0)}{\omega_c(0)} \,\, ; \,\,
    C_2 = \frac{- 2\pi \tilde{A} B}{n_H - n_L} \,,
\end{equation}
with 
\begin{equation}
    \tilde{A} = n_H n_L A_L = \frac{1}{2}(1 + \frac{n_L}{n_H} + \frac{n_H}{n_L} - n_H n_L) \,.
\end{equation}
For the SiO$_2$ / Ta$_2$O$_5$ example discussed above, the L-DBR is expected to show a smaller Bragg effect as it has $\tilde{A} = 0.08$, while the H-DBR has $A=0.66$.  



\section{Operator algebra}
\label{sec:Appendix-operators}

This appendix introduces ladder operators for the scalar LG-modes and shows how they can be used to calculate the matrix elements of the perturbing operators. 
It also shows how these concepts can be applied to vector LG-modes, including the $X/Y/A/B$ modes introduced in the main text.
We will only consider coupling between modes with the same transverse order $N$, such that the wave-front curvature and Gouy phase drop out of the problem.     

In Cartesian $(x,y)$ coordinates, with normalized coordinates $(\xi,\eta)=(x,y)/\gamma_z$, we define the creation and annihilation operators in the $\xi$ direction as \cite{Nienhuis}
\begin{equation}
    \hat{a}_\xi(\chi) = \frac{1}{\sqrt{2}} (\xi + \partial_\xi) \quad ; \quad \hat{a}^\dagger_\xi(\chi) = \frac{1}{\sqrt{2}} (\xi - \partial_\xi)  
\end{equation}
and likewise for the $\eta$ direction. 
These ladder operators allow one to ladder through the set of scalar HG-modes. 
As our system is approximately rotational symmetric, it is more convenient to work with cylindrical coordinates $(\rho,\theta)$ and the circular ladder operators
\begin{equation}
    \hat{a}_\pm = \frac{1}{\sqrt{2}} (\hat{a}_\xi \mp i \hat{a}_\eta) \quad ; \quad \hat{a}^\dagger_\pm = \frac{1}{\sqrt{2}} (\hat{a}^\dagger_\xi \pm i \hat{a}^\dagger_\eta) \,.
\end{equation}
These ladder operators satisfy the commutation relation $[\hat{a}_i , \hat{a}_j^\dagger ] = \delta_{ij}$ and combine into number operators $\hat{n}_+ = \hat{a}_+^\dagger \hat{a}_+$ and $\hat{n}_- = \hat{a}_-^\dagger \hat{a}_-$. 
They allow one to ladder through the set of scalar LG-modes, using the relation 
\begin{eqnarray}
\label{eq:scalar-LG}
   \sqrt{n_+! n_-!}  | \tilde{\Psi}_{n_+,n_-} \rangle & = & \left( \hat{a}_+^\dagger \right)^{n_+} \left( \hat{a}_-^\dagger \right)^{n_-} | \tilde{\Psi}_{0,0} \rangle \,, \\
    \langle \xi,\eta | \tilde{\Psi}_{n_+,n_-} \rangle & = & e^{i(n_+-n_-) \theta} f_{p\ell}(\rho) \,,
\end{eqnarray}
where $| \tilde{\Psi}_{0,0} \rangle$ is the fundamental mode. 
The final equation provides the link to the modes used in the main text. 
We have added a tilde to the notation to indicate that these LG-modes are labeled with quantum numbers $n_+$ and $n_-$. 
The relation with the quantum numbers used in the main text is $p = {\rm min}(n_+,n_-)$ and $\ell = n_+ - n_-$, where the later can still be positive or negative.

Next, we will express each perturbing operator as combinations of ladder operators and calculate the matrix representation of these operators in the basis of the LG-modes of transverse order $N$.
Using the ladder operators introduced above, we can for instance rewrite the paraxial form $\rho^2 = \xi^2 + \eta^2$ and $\Delta_\perp = \partial_\xi^2 + \partial_\eta^2$ as
\begin{eqnarray}
    \hat{\rho}^2 & = & (\hat{N}+1) + \hat{a}_+ \hat{a}_- + \hat{a}^\dagger_+ \hat{a}^\dagger_- \,, \\
    \hat{\Delta}_\perp & = & - (\hat{N}+1) + \hat{a}_+ \hat{a}_- + \hat{a}^\dagger_+ \hat{a}^\dagger_- \,, 
\end{eqnarray}
where $\hat{N} = \hat{n}_+ + \hat{n}_-$.
When we sandwich these operators between two LG-modes of order $N$, the first term yields the familiar expression $\langle \tilde{\Psi}_{j'} | \hat{\rho}^2 | \tilde{\Psi}_j \rangle = (N+1) \delta_{j'j}$, while the second and third term do not contribute as they only couple modes with different order. 

The quartic non-paraxial operators yield expressions with more terms.
When we only keep the operator combinations that couple modes of the same order $N$, we find that the quartic non-paraxial operators are diagonal in the LG-basis with
\begin{equation}
    \langle \tilde{\Psi}_j | \hat{\rho}^4 | \tilde{\Psi}_j \rangle = (N+1)^2 + N+1 + 2n_+n_- = f(p,\ell) \,,
\end{equation}
where $j = (n_+, n_-)$, and an identical result for $\langle \tilde{\Psi}_j | \hat{\Delta}_\perp^2 | \tilde{\Psi}_j \rangle$. 
The final expression shows the link to the quadratic polynomial $f(p,\ell)$, used in the main text. 
The astigmatic operator 
$\hat{\xi}^2-\hat{\eta}^2 = \hat{a}^\dagger_+ \hat{a}_- + \hat{a}^\dagger_- \hat{a}_+$
simultaneously lowers $n_+$ and raises $n_-$ by one, or visa versa, and can therefore couple modes of the same order $N$ with $\Delta \ell = \pm 2$. 
More precisely
\begin{eqnarray}
\label{eq:astigmatism}
    (\hat{\xi}^2-\hat{\eta}^2) | \tilde{\Psi}_{n_+,n_-} \rangle & = & 
    \sqrt{(n_++1)n_-} | \tilde{\Psi}_{n_++1,n_--1} \rangle \nonumber \\
 & + & \sqrt{n_+(n_-+1)} | \tilde{\Psi}_{n_+-1,n_++1} \rangle \,.
\end{eqnarray}
When we label the scalar modes with a single quantum number $n_s=n_+$, such that $n_-=N-n_s$, the only non-zero elements of the astigmatic operator are
\begin{equation}
    (\hat{\xi}^2-\hat{\eta}^2)_{n_s+1,n_s} =  (\hat{\xi}^2-\hat{\eta}^2)_{n_s,n_s+1} = h(N,n_s) \,,
\end{equation}
where the function $h(N,n_s) \equiv \sqrt{(n_s+1)(N-n_s)}$ obeys the symmetry $h(N,N-n_s)=h(N,n_s-1)$.

Next, we introduce the vector modes via
\begin{equation}
\label{eq:vector-mode1}
    | \vec{\Psi} \rangle = \vec{e}_+ | \Psi_+ \rangle + \vec{e}_- | \Psi_- \rangle  \,,
\end{equation}
where $\vec{e}_\pm = (\vec{e}_x \pm i \vec{e}_y)/\sqrt{2}$ are the circular-polarized unit vectors and where $| \Psi_\pm \rangle$ are scalar mode profiles. 
We write this vector mode as a vector of two scalar modes and describe the action of any tensor operator ${\cal H}$ by the associated $2\times2$ tensor that acts via
\begin{equation}
\label{eq:vector-mode2}
\begin{pmatrix} {\cal H}_{++} & {\cal H}_{+-}  \\ {\cal H}_{-+}  & {\cal H}_{--} \end{pmatrix} \begin{pmatrix} | \Psi_+ \rangle \\ | \Psi_- \rangle \end{pmatrix} \,.
\end{equation}

We will consider three vector corrections: (i) the isotropic spin-orbit coupling, (ii) the anisotropic spin-orbit coupling, and (iii) the Bragg effect. 
The most prominent vector correction is the isotropic spin-orbit coupling described by the operator ${\cal H}_{\rm vec} = 2/(kR_m) \, \vec{\rho} \otimes \vec{\nabla}_\rho$.
The $2 \times 2$ matrix representation of the operator $\vec{\rho} \otimes \vec{\nabla}_\rho$ contains the combinations $\{\hat{\xi} \partial_\xi , \hat{\xi} \partial_\eta , \hat{\eta} \partial_\xi  , \hat{\eta} \partial_\eta \}$ in the linear-polarized $(\xi,\eta)$ basis. 
Conversion of these expression to the circular ladder operators and to the circular-polarized form defined in Eqs. (\ref{eq:vector-mode1}) and (\ref{eq:vector-mode2}) yields
\begin{equation}
{\cal H}_{\rm vec} = \frac{-1}{kR_m} \begin{pmatrix} 1 + n_+ - n_-  & 0  \\ 0 & 1 - n_+ + n_- \end{pmatrix} \,.
\end{equation}
In this conversion, we removed combinations of operators that only projects to modes of different order $N$, like the operator $\hat{\xi} \partial_\xi + 1 = \frac{1}{2}[\hat{a}_\xi^2 - (\hat{a}^\dagger_\xi)^2]$ and combinations of the form $\hat{a}_\xi \hat{a}_\eta$ and $\hat{a}^\dagger_\xi \hat{a}^\dagger_\eta$.

Application of ${\cal H}_{\rm vec}$ to the righthand circular-polarized modes yields
\begin{equation}
\label{eq:H_vec}
    {\cal H}_{\rm vec} \, \vec{e}_+ |\tilde{\Psi}_{n_+,n_-} \rangle = - \frac{1+\tilde{\ell}}{kR_m} \,\, \vec{e}_+ |\tilde{\Psi}_{n_+,n_-} \rangle \,,
\end{equation}
where $\tilde{\ell} = n_+-n_-$, with associated matrix elements 
\begin{eqnarray}
\label{eq:H_vec2}
\left( \left( {\cal H}_{\rm vec} \right)_{++}\right)_{n_s,n_s} =  - \frac{1+\tilde{\ell}}{kR_m} = - \frac{1+2n_s-N}{kR_m} \,.
\end{eqnarray}
The middle part of Eq. (\ref{eq:H_vec2}) is identical to Eq.~(\ref{eq:vec}) in the main text, as the signed $\tilde{\ell} = \pm \ell$ depending $n_+ \lessgtr n_-$.
The right side of Eq. (\ref{eq:H_vec2}) introduces $n_s$ as the circular quantum number along the spin direction, such that $n_s = n_+$ for $s=1$ and $n_s = n_-$ for $s=-1$.
With these definitions, the only non-zero matrix elements of $\left( {\cal H}_{\rm vec} \right)_{--}$ are
\begin{eqnarray}
\label{eq:H_vec3}
\left( \left( {\cal H}_{\rm vec} \right)_{--}\right)_{n_s,n_s} =  - \frac{1-\tilde{\ell}}{kR_m} = - \frac{1+2n_s-N}{kR_m} \,.
\end{eqnarray}

The anisotropic component of the spin-orbit coupling follows from the Taylor expansion
\begin{eqnarray}
    (\frac{x \vec{e}_x }{kR_x} + \frac{y \vec{e}_y}{kR_y} )  \otimes \vec{\nabla}_\perp & \approx &  (\frac{x \vec{e}_x + y \vec{e}_y }{k \overline{R}})  \otimes \vec{\nabla}_\perp \\ 
    & + & \eta_{\rm astigm} (\frac{x \vec{e}_x - y \vec{e}_y }{k \overline{R}})  \otimes \vec{\nabla}_\perp \,, \nonumber
\end{eqnarray}
where $\overline{R} = (R_x + R_y)/2$, $\eta_{\rm astigm} = (R_y - R_x)/\overline{R}$.
The first operator on the right-hand side is ${\cal H}_{\rm vec}/2$. 
The second operator describes the anisotropic spin-orbit coupling ${\cal H}_{\rm v+a}/2$. 
The $2 \times 2$ matrix representation of this operator in $(\xi,\eta)$ coordinates and polarization contains the combinations $\{\hat{\xi} \partial_\xi , \hat{\xi} \partial_\eta , - \hat{\eta} \partial_\xi  , - \hat{\eta} \partial_\eta \}$. 
Conversion to the circular-polarized vector basis yields
\begin{equation}
{\cal H}_{\rm v+a} = \frac{-\eta_{\rm astigm}}{kR_m} \begin{pmatrix} 0 & 1 - n_+ + n_-   \\ 1 + n_+ - n_- & 0 \end{pmatrix} \,.
\end{equation}
The off-diagonal elements show how this operator converts $s=+1 \Leftrightarrow s=-1$ circular-polarized light.
Application of ${\cal H}_{\rm v+a}$ to the righthand circularly-polarized modes yields
\begin{equation}
    {\cal H}_{\rm v+a} \, \vec{e}_+ |\tilde{\Psi}_{n_+,n_-} \rangle = - \frac{\eta_{\rm astigm}}{kR_m}\,(1+\tilde{\ell}) \,\, \vec{e}_- |\tilde{\Psi}_{n_+,n_-} \rangle \,.
\end{equation}
This equation differs in two ways from Eq. (\ref{eq:H_vec}).
First of all, the extra factor $\eta_{\rm astigm}$ shows that anisotropic spin-orbit coupling is linked to astigmatism.
Second, the ${\cal H}_{\rm v+a}$ operator changes the handedness of the circular polarization.
As a result the projected circular quantum number changes from $n_{s,in}=n_+$ to $n_{s,out}=n_-=N-n_s$.
The associated matrix in the circular-polarized basis therefore only has anti-diagonal elements
\begin{equation}
\left( \left( {\cal H}_{\rm v+a} \right)_{-+} \right)_{N-n_s,n_s} =  - \frac{\eta_{\rm astigm}}{kR_m}\, (1+2n_s-N) \,, 
\end{equation}
where the first mode label refers to $n_- = N-n_s$ and the second mode label refers to $n_+ = n_s$. A similar analysis for the lefthand circular polarized modes yields the identical, Hermitian-conjugated, result
\begin{equation}
\left( \left( {\cal H}_{\rm v+a} \right)_{+-} \right)_{N-n_s,n_s} =  - \frac{\eta_{\rm astigm}}{kR_m}\, (1+2n_s-N) \,. 
\end{equation}

The final vector correction originates from the Bragg effect described by Eq.~(\ref{eq: hamiltonian bragg effect}). 
We will only consider the $k_\perp^2$ contribution to the Bragg effect and neglect the $k_\perp^4$ contribution, which is typically weaker but can still be relevant at small frequency detuning.
In the linearly-polarized basis and $x,y$ units used in the main text, this part of the Bragg operator has the form
\begin{equation}
{\cal H}_{\rm Bragg} = \frac{C_0}{k^2} \begin{pmatrix} k_x^2-k_y^2 & 2k_xk_y \\ 2k_xk_y & k_y^2 - k_x^2 \end{pmatrix} \,,   
\end{equation}
where $k_x = i\partial_x$ and $k_y = i\partial_y$.
Conversion to normalized coordinates and to the preferred circular-polarized vector basis yields
\begin{equation}
{\cal H}_{\rm Bragg} = \frac{2C_0}{kz_0} \begin{pmatrix} 0 & \hat{a}^\dagger_- \hat{a}_+ \\ \hat{a}^\dagger_+ \hat{a}_- & 0 \end{pmatrix} \,.   
\end{equation}
The off-diagonal elements show how this operator also converts $s=+1 \leftrightarrow s=-1$ circular-polarized light.
It does so under conservation of total angular momentum $J = l + s$, such that $\Delta \ell = \Delta n_+ - \Delta n_- = \pm2$.

Application of ${\cal H}_{\rm Bragg}$ to the righthand circular-polarized modes yields
\begin{equation}
    {\cal H}_{\rm Bragg} \, \vec{e}_+ |\tilde{\Psi}_{n_+,n_-} \rangle = \frac{2C_0}{kz_0} \, h(N,n_s) \,\, \vec{e}_- |\tilde{\Psi}_{n_++1,n_--1} \rangle \,.
\end{equation}
where $h(N,n_s) = \sqrt{(n_s+1)(N-n_s)}$ as before.
The projected circular quantum number now changes from $n_{s,in}=n_+$ to $n_{s,out}=n_--1=N-n_s-1$ and the associated matrix only has elements one row below the anti-diagonal, with
\begin{eqnarray}
\left( \left( {\cal H}_{\rm Bragg} \right)_{-+} \right)_{N-n_s-1,n_s} =  \frac{2C_0}{kz_0}\, h(N,n_s) \,.
\end{eqnarray}
A similar analysis for the lefthand circular polarized modes again yields identical matrix elements for $\left( {\cal H}_{\rm Bragg} \right)_{+-}$, due to our use of projected indices.

\section{Hyperfine splittings}
\label{sec:Appendix-hyperfine}

Appendix \ref{sec:Appendix-operators} showed that two vector corrections can change the handedness of the light, i.e. have non-zero operators ${\cal H}_{+-}$ and ${\cal H}_{-+}$.
As a result, the circular polarized modes become coupled.
In this Appendix we will argue that the vector LG-modes introduced in the main text are the new eigenmodes of this coupled system. 
The coupling can lift the original two-fold degeneracy of some vector LG-modes, though, and create a hyperfine splitting between the $+$ and $-$ versions of some $\ell A$ or $\ell B$ modes.

In the main text we introduced a special set of vector LG-modes that we labeled by their absolute OAM $\ell \geq 0$, their X/Y/A/B character, and their $\pm$ polarity under $x$-mirror reflection.
The link between these vector LG-modes and the scalar LG-modes in Eq. (\ref{eq:scalar-LG}) is 
\begin{eqnarray}
\label{eq:vectorLGA+}
| \ell A + \rangle & = & \sqrt{2} \, {\rm Re} [\, \vec{e}_+ |\tilde{\Psi}_{\frac{N-\ell}{2},\frac{N+\ell}{2}} \rangle \,] \,, \\
| \ell A - \rangle & = & \sqrt{2} \, {\rm Im} [\, \vec{e}_+ |\tilde{\Psi}_{\frac{N-\ell}{2},\frac{N+\ell}{2}} \rangle \,] \,, \\
| \ell B + \rangle & = & \sqrt{2} \, {\rm Re} [\, \vec{e}_+ |\tilde{\Psi}_{\frac{N+\ell}{2},\frac{N-\ell}{2}} \rangle \, ] \,, \\
\label{eq:vectorLGB-}
| \ell B - \rangle & = & \sqrt{2} \, {\rm Im} [\, \vec{e}_+ |\tilde{\Psi}_{\frac{N+\ell}{2},\frac{N-\ell}{2}} \rangle \,]  \,,
\end{eqnarray}
where ${\rm Re}$ and ${\rm Im}$ denote the real- and imaginary part, with $\vec{e}_+^{\, *} = \vec{e}_-$ and  
 $|\tilde{\Psi}_{\frac{N-\ell}{2},\frac{N+\ell}{2}} \rangle ^* = |\tilde{\Psi}_{\frac{N+\ell}{2},\frac{N-\ell}{2}} \rangle $.
These vector LG-modes are the true eigenmodes of the perturbed cavity, as the two vector corrections that we consider are symmetric under $x$-mirror reflection and hence cannot couple $+$ and $-$ modes. 
Equations (\ref{eq:vectorLGA+})-(\ref{eq:vectorLGB-}) show that A modes are like B modes with signed OAM $\tilde{\ell} = -\ell$ instead of $\tilde{\ell} = \ell$ in the $\vec{e}_+$ component of their vector field. 
The $\ell =0$ modes obey the relations $|0X+\rangle = |0A+\rangle = |0B+\rangle$ and  $|0Y-\rangle = |0A-\rangle = |0B-\rangle$.

Application of ${\cal H}_{v+a}$ to the vector LG-modes yields 
\begin{eqnarray}
\label{eq:Hva1}
{\cal H}_{\rm v+a} \, | \ell A \pm \rangle & = & \pm \frac{\eta_{\rm astigm}}{kR_m}\,(\ell-1) \,\, | \ell B \pm \rangle \,, \\
\label{eq:Hva2}
{\cal H}_{\rm v+a} \, | \ell B \pm \rangle & = & \pm \frac{\eta_{\rm astigm}}{kR_m} \, (-\ell-1) \,\, | \ell A \pm \rangle \,.
\end{eqnarray}
For $\ell = 0$, the ${\cal H}_{\rm v+a}$ operator has on-diagonal element $-\eta_{\rm astigm}/(kR_m)$ for the $|0X+\rangle$ modes and $\eta_{\rm astigm}/(kR_m)$ for the $|0Y- \rangle$ mode. 
This difference creates the hyperfine splitting of the $\ell =0$ modes described in the main text.
For $\ell \geq 1$, the ${\cal H}_{\rm v+a}$ operator has off-diagonal elements that couple $|\ell A \pm \rangle$ to $|\ell B \pm \rangle$ in an asymmetric way, related to the $\pm \ell$ alignment of the OAM. 

Application of ${\cal H}_{\rm Bragg}$ to the vector LG-modes yields 
\begin{eqnarray}
\label{eq:Bragg-B}
{\cal H}_{\rm Bragg} | \ell B \pm \rangle & = & \pm \frac{2C_0}{kz_0} h(N,\frac{N+\ell}{2}) \,\, | (\ell+2) A \pm \rangle \,, \\
\label{eq:Bragg-A}
{\cal H}_{\rm Bragg} | \ell A \pm \rangle & = & \pm \frac{2C_0}{kz_0} h(N,\frac{N-\ell}{2}) \,\, | (\ell-2) B \pm \rangle \,.
\end{eqnarray}
The $\ell=1$ case of Eq. (\ref{eq:Bragg-A}) corresponds to ${\cal H}_{\rm Bragg} |\ell A \pm\rangle = \pm C_0 (N+1)/(kz_0) |\ell A \pm\rangle$, if we interpret the mode $|-1,B,\pm \rangle = |1 A \pm \rangle$. 
This creates the hyperfine splitting between the $1A+$ and $1A-$ mode described in the main text.  
The $\ell=0$ case of Eq. (\ref{eq:Bragg-A}) is effectively described by Eq. (\ref{eq:Bragg-B}), as $|0X+\rangle = |0A+\rangle = |0B+\rangle$.

To visualize the obtained results, we end by deriving the full ${\cal H}_{++}$ and ${\cal H}_{--}$ matrices in the vector LG-mode basis for the $N=1$ and $N=2$ subspace.
We quantify the relative strength of the astigmatism with the parameter $X=(8\pi k R_m) \eta_{\rm astigm} \tan{\chi_0}/(2\pi)$ used in Sec. \ref{sec:astigm} and add the two polarization-changing effects mentioned above. 
In the normalized units used below, the on-diagonal elements of the spin-orbit coupling are $-4(\tilde{\ell}+1)$, where $\tilde{\ell}=\ell$ for $B$ modes and $\tilde{\ell}=-\ell$ for $A$ modes. 
In the same units, the anti-diagonal elements associated with the anisotropic spin-orbit coupling are $\mp 4 (\tilde{\ell}+1) \eta_{\rm astigm}$, see Eqs. (\ref{eq:Hva1}) and (\ref{eq:Hva2}).
In the same units, the matrix elements of the Bragg correction are of the form $\pm Y h(N,n_s)$, see Eqs. (\ref{eq:Bragg-B}) and ({eq:Bragg-A}). 
Combination of these contributions for the $N=1$ group changes Eq. (\ref{eq:N1}) into 
\begin{equation}
    (8\pi k R_m)\, \Delta \tilde{\nu}_{(N=1)} = \begin{pmatrix} -6 & X \\ X \mp 8\eta_{\rm astigm} & 2 \pm Y \end{pmatrix} 
\end{equation}
where the upper signs describe the spectral matrix of the $(1B+,1A+)$ modes and the lower signs that of the $(1B-,1A-)$ modes.
The $\pm Y$ on-diagonal element describes the hyperfine splitting of the $1A\pm$ modes.
The $\mp 8 \eta_{\rm astigm}$ off-diagonal element can also produce some hyperfine splitting in strongly astigmatic cavities, but this effect is typically very small as $\eta_{\rm astigm} \ll 1$.

For the $N=2$ group the hyperfine splittings change Eq. (\ref{eq:N2}) into
\begin{widetext}
 \begin{equation}
    (8\pi k R_m)\, \Delta \tilde{\nu}_{(N=2)} = \begin{pmatrix} -12 & \sqrt{2}X & \pm 4\eta_{\rm astigm} \\ 
    \sqrt{2}X & 2 \mp 4\eta_{\rm astigm}  & \sqrt{2}X \pm \sqrt{2}Y \\ \mp 12 \eta_{\rm astigm}  & \sqrt{2}X \pm\sqrt{2}Y & 4  \end{pmatrix} \,.
\end{equation}   
\end{widetext}
where the upper/lower signs refer to the $(2B+,0X+,2A+)$ and $(2B-,0Y-,2A-)$ basis, respectively.
The $\mp 4 \eta_{\rm astigm}$ on-diagonal element describes the (typically small) hyperfine splitting between the $0X$ and $0Y$ due to shape birefringence.
The $\pm \sqrt{2} Y$ off-diagonal elements are typically also small and will only produce a measurable hyperfine splitting in strongly astigmatic cavities, where the mode mixing induced by the off-diagonal $\sqrt{2}X$ element makes the new eigenmodes sensitive to the Bragg effect. 
The spectral matrix shows that the resulting hyperfine splittings are stronger for the $0$ and $2A$ modes than for the $2B$ modes. 


\bibliographystyle{apsrev4-1}
\bibliography{references2(apsstyle)}
\end{document}